%% file: paper.tex
\documentclass[acmsmall]{acmart}
\AtBeginDocument{%
  \providecommand\BibTeX{{%
    \normalfont B\kern-0.5em{\scshape i\kern-0.25em b}\kern-0.8em\TeX}}}

\setcopyright{acmcopyright}
\copyrightyear{2023}
\acmYear{2023}
\acmDOI{XXXXXXX.XXXXXXX}

\acmJournal{TOSEM}
\acmVolume{0}
\acmNumber{0}
\acmArticle{0}
\acmMonth{0}

\usepackage{multirow}
\usepackage{color,soul}
\usepackage{framed}

\newcommand{\rev}[1]{{\color{black}#1}}
\newcommand{\added}[1]{{\color{black}#1}}

\begin{document}

\title{Using Voice and Biofeedback to Predict User Engagement
during Product Feedback Interviews}


\author{Alessio Ferrari}
\email{alessio.ferrari@isti.cnr.it}
\orcid{0000-0002-0636-5663}
\affiliation{%
  \institution{Consiglio Nazionale delle Ricerche, Istituto di Scienza e Tecnologie della Comunicazione ``A. Faedo'' (CNR-ISTI)}
  \city{Pisa}
  \country{Italy*}
}

\author{Thaide~Huichapa}
\email{thuichap@students.kennesaw.edu}
\affiliation{%
  \institution{Kennesaw State University (KSU)}
  \city{Marietta Campus}
  \state{Georgia}
  \country{U.S.A.}
}

\author{Paola~Spoletini}
\email{pspoleti@kennesaw.edu}
\orcid{0000-0001-7922-4936}
\affiliation{%
  \institution{Kennesaw State University (KSU)}
  \city{Marietta Campus}
  \state{Georgia}
  \country{U.S.A.}
}

\author{Nicole Novielli}
\email{nicole.novielli@uniba.it}
\orcid{0000-0003-1160-2608}
\affiliation{%
  \institution{University of Bari}
  \city{Bari}
  \country{Italy}
}

\author{Davide Fucci}
\email{davide.fucci@bth.se}
\orcid{0000-0002-0679-4361}
\affiliation{%
  \institution{Blekinge Tekniska Högskola (BTH)}
  \city{Karlskrona}
  \country{Sweden}
}

\author{Daniela Girardi}
\email{daniela.girardi@uniba.it}
\orcid{0000-0002-4630-4793}
\affiliation{%
  \institution{University of Bari}
  \city{Bari}
  \country{Italy}
}

\renewcommand{\shortauthors}{Ferrari, et al.}

\begin{abstract}
  Capturing users' engagement is crucial for gathering feedback about the features of a software product.
In a market-driven context, current approaches to collecting and analyzing users' feedback are based on techniques leveraging information extracted from 
product reviews and social media.   
These approaches are hardly applicable \rev{in contexts where online feedback is limited, as for the majority of apps, and software in general.}
In such cases, companies need to resort to face-to-face interviews to get feedback on their products. 
In this paper, we propose to utilize biometric data, in terms of physiological and voice features, to complement product feedback interviews with information about the engagement of the user on product-relevant topics. 
We evaluate our approach by interviewing users while gathering their physiological data (i.e., \textit{biofeedback}) using an Empatica E4 wristband, and capturing their voice through the default audio-recorder of a common laptop. 
Our results show that we can predict users' engagement by training supervised machine learning algorithms on biofeedback and voice data, and that voice features alone can be sufficiently effective. \added{The best configurations evaluated achieve an average F1 $\sim 70\%$ in terms of classification performance, and use voice features only}. This work is one of the first studies in requirements engineering in which biometrics are used to identify emotions. Furthermore, this is one of the first studies in software engineering that considers voice analysis. 
The usage of voice features can be particularly helpful for emotion-aware \rev{feedback collection} in remote communication, either performed by human analysts or voice-based chatbots, and can also be exploited to support the analysis of meetings in software engineering research.
\end{abstract}

\ccsdesc[500]{Software and its engineering~Requirements analysis}
\ccsdesc[500]{Information systems~Information systems applications}

\keywords{requirements engineering, emotion detection, voice analysis, biofeedback analysis}

\received{Day Month Year}

\maketitle

\input{section/introduction}
\input{section/background}
\input{section/design}

\input{section/results}

\input{section/discussion}

\input{section/threats}
\input{section/conclusion}

\begin{acks}
This study was carried out within the MOST – Sustainable Mobility National Research Center and received funding from the European Union Next-GenerationEU (PIANO NAZIONALE DI RIPRESA E RESILIENZA (PNRR) – MISSIONE 4 COMPONENTE 2, INVESTIMENTO 1.4 – D.D. 1033 17/06/2022, CN00000023). 
This work was also partially supported by National MIUR-PRIN 2020TL3X8X project T-LADIES (Typeful Language Adaptation for Dynamic, Interacting and Evolving Systems), by the National Science Foundation under grant CCF-1718377, European Union’s Horizon Europe research and innovation programme under grant agreement no. 101060179 (CODECS).
\end{acks}

\bibliographystyle{ACM-Reference-Format}
\bibliography{bibliography}


\section{Online Resources}
A replication package is shared at~\cite{alessio2021package}.
\end{document}

%% file: section/introduction.tex
\section{Introduction}\label{sec:introduction}

The development of novel software products, as well as the improvement of existing ones, can deeply benefit from the involvement of users in requirements  engineering (RE) activities~\cite{bano2015systematic}. Getting feedback from the user base has been recognised to lead to increased usability, improved satisfaction~\cite{bakalova2011comparative},  better understanding of requirements~\cite{hanssen2006agile}, and creation of long-term relationships with customers~\cite{heiskari2009investigating}. 

User feedback can take implicit and explicit forms, and different means are available to collect this information. In particular, data analytics applied to user opinions and to usage data has seen an increasing interest in the last years, leading to the birth of RE sub-fields such as \textit{crowd RE}~\cite{murukannaiah2016acquiring,groen2017crowd} and \textit{data-driven RE}~\cite{maalej2015toward,williams2017mining}. \rev{This is particularly common in the field of mobile app development, where user reviews are normally exploited to drive RE and other software engineering activities, as shown in the recent survey by Al-Subaihin \textit{et al.}~\cite{al2019app}. However Dabrowski \textit{et al.}~\cite{dkabrowski2022analysing} also observed that while developers of apps with a large user base can profit from data analytics tool, the majority of apps receive a limited number of reviews, and it is unclear whether these small businesses can fully profit from online feedback analysis. In these cases, companies need to complement the information with more traditional RE techniques, such as prototyping, observations, usability testing, and focus groups~\cite{zowghi2005requirements}}.
Among the classical RE techniques, user interviews are one of the most commonly used \added{techniques} to gather requirements and feedback~\cite{fernandez2017naming,davis2006effectiveness,hadar2014role,palomares2021state}. \rev{User interviews are also common in the field user experience (UX)~\cite{wilson2013interview}.} 
Several aspects have been observed to influence the success and failure of interviews, such as the domain knowledge of the \rev{interviewers}~\cite{hadar2014role,aranda2015effect}, ambiguity in communication~\cite{ferrari2016ambiguity}, and typical mistakes such as not providing a wrap-up summary at the end of the interview session, or not creating rapport with the interviewee~\cite{bano2019teaching}. 

The field of \textit{affective RE}---which combines ideas from Affective Computing with RE practices---recognised the role of users' emotions, and different studies address emotion-related problems in RE.
Contributions include applications of sentiment analysis to app reviews~\cite{GM14,KM18}, analysis of users' facial expressions~\cite{SKE19,MSE19}, the study of physiological reactions to ambiguity~\cite{spoletini2016empowering}, and the augmentation of goal models with user emotions elicited through psychometric surveys~\cite{TSP19}. Currently, little attention is dedicated to the emotional aspects of interviews and, in particular, to users' \textit{engagement}.
Capturing engagement, \added{and emotions in general}, is crucial for gathering feedback about the features of a certain product, and having a better understanding of the users' preferences\added{~\cite{saariluoma2014emotional,sutcliffe2022designing}}. \added{This can facilitate the prioritisation of requirements towards product evolution~\cite{achimugu2014systematic}, and optimise release planning~\cite{nayebi2018asymmetric,ruhe2010product,villarroel2016release}}. 
\added{For example, if requirements analysts are able to understand which existing or planned features of a software application have the potential to create more positive engagement with users, they can better prioritise further development of such features, or make them more easily accessible from the user interface. Instead, if other features are considered annoying, they can identify ways to improve them or decide to discard them from the next release. Although prioritisation information can be identified, e.g., with questionnaires, interviews can capture more in-depth feedback and nuanced opinions on product-related aspects. 
} 



In this paper, we aim to extend the body of knowledge in affective RE by studying users' emotions during \rev{software product feedback interviews}. 
\added{We focus on \textit{engagement}---i.e., an affective state indicating the degree of positive or negative emotional involvement in a certain product-related aspect discussed in the interview.
We hypothesize that different forms of engagement will arise based on the emotional context. Positive-high engagement is expected when users discuss topics they like and feel strongly about. Conversely, negative-high engagement may occur when users engage with contentious subjects they dislike. Low engagement, on the other hand, is anticipated when users lack strong feelings, leading to either a calm or disinterested state. 
} 

We perform a study with 31 participants taking part in a simulated interview during which we capture their biofeedback using an Empatica E4 wristband, we record  their voice through a common laptop recorder, and collect their self-assessed engagement. \rev{The interview is designed to mimic a context in which the user provides verbal feedback about a software product.}  
We compare different machine learning algorithms to predict users' engagement based on features extracted from biofeedback and voice signals. \added{The idea is that, by automatically detecting engagement, prioritisation of requirements can be carried out without explicitly asking about preferences, thus making the feedback interview more natural. Furthermore, during the interview, one can identify the engagement associated not only with features, but also with high-level product-related aspects (e.g., ethics, privacy), for which it is hard to articulate a clear-cut preference or priority.}


Our experiments show that engagement can be predicted in terms of valence and arousal~\cite{Russel1991} with an F1-measure of, respectively, 63\% and 65\%, when only considering biofeedback signals. When using voice signals alone, the performance in terms of F1-measure increases to \added{68\% and 71\%}, showing that voice features can be strongly predictive of users' engagement. \added{The combination of biofeedback and voice features does not increase the performance compared to voice feature alone---F1-measure of 68\% and 69\%}.


This paper builds upon a previous conference contribution by the same authors~\cite{girardi2020theway}, in which only the biofeedback signals were used for prediction. The current paper replicates and expands the experiments. In particular, we introduce additional biometric features, based on voice signals, as well as additional data preparation options---namely standard scaling, oversampling and data imputation---which allow us to improve the previous results, even using only voice features.
This is a particularly relevant outcome, as voice analysis can be scaled, with minimal costs, to a large number of interviewees, who may be located remotely. The idea could be effective in case voice-based conversational agents are used as interviewers, 
to equip them with the ability to detect the engagement of the user from their emotional prosody~\cite{buchanan2000recognition}, and adapt the interview accordingly, \added{for example by asking more in-depth questions about the discussion topics that appear to create more engagement. The degree of engagement identified through the analysis of voice can also help to perform automatic prioritisation of requirements.} \added{This is expected to be feasible in the near future, thanks to the recent development in the use of chatbots for product feedback analysis~\cite{wolfinger2022chatbot}, and the potential offered by disruptive technologies for conversation/question-answering such as ChatGPT\footnote{\url{https://openai.com/blog/chatgpt}.  Visited 18 Sept. 2023.}, which recently supports also speech-based interaction\footnote{\url{https://openai.com/blog/chatgpt-can-now-see-hear-and-speak}. Visited 27 Sept. 2023.}.}


This paper contributes to the literature with empirical work in which we show that  biometric features, including physiology- and voice-related metrics, can be applied to predict users' engagement during requirements interviews. While biofeedback analysis has been previously applied in other software engineering areas~\cite{Girardi2020,FritzBMYZ14,girardi2021emotions}, this is one of the first works that explores its potential in RE. Furthermore, to our knowledge, this is the first work that uses voice analysis in software engineering, thus opening to studies focused on speech-intensive activities of software development. 
A replication package is also made available\footnote{https://doi.org/10.5281/zenodo.8383385}~\cite{alessio2021package} to enable other researchers to build on our results.

The remainder of the paper is structured as follows.
In Section~\ref{sec:background}, we present background definitions of engagement and theories of emotions, as well as related work in RE and software engineering.
In Section~\ref{sec:design}, we report our study design, whereas Section~\ref{sec:results} reports its results.
We discuss the implications of our study in Section~\ref{sec:discussion} and its limitations in  Section~\ref{sec:threats}. 
Finally, Section~\ref{sec:conclusion} concludes the paper.

%% file: section/background.tex
\section{Background and Related Work}
\label{sec:background}
In this section, we first clarify the relationship between emotion modelling and engagement (Sect.~\ref{sec:engagementandemotions}). Then, we present the background on affect modelling and emotion classification using biofeedback (Sect.~\ref{sec:biofeedbackbasedclassification}) and voice analysis (Sect.~\ref{sec:voiceanalysisclassification}). Finally, we discuss relevant related work in RE and software engineering
(Sect.~\ref{sec:biofeedbackvoiceinSE}).

\subsection{Engagement and Emotions}
\label{sec:engagementandemotions}

Affective states vary in their degree of stability, ranging from personality traits---i.e., long-standing, organized sets of characteristics of a person---to emotions---i.e., transient and typically episodic, dynamic, and structured events. 
Emotions involve perceptions, thoughts, feelings, bodily changes, and personal dispositions to experience them.
Emotions are episodic and dynamic in that, over time, they can vary depending on several factors~\cite{cowie2011emotion}. 

Psychologists have investigated the nature and triggers of emotions for decades, leading to a plethora of theories of emotions. Among these theories, cognitive models describe emotions as reactions to cognition. For example, the OCC model~\cite{OCC:1988} defines a taxonomy of emotions and identifies them as \textit{valenced} reactions (either positive or negative) to the cognitive processes involved in evaluating objects, events, and agents.
Analogously, Lazarus describes nine negative (Anger, Anxiety, Guilt, Shame, Sadness, Envy, Jealousy, and Disgust) and six positive (Joy, Pride, Love, Relief, Hope, and Compassion) emotions, as well as their \textit{appraisal} patterns:  when a situation is congruent with the person's goals positive emotions arise; otherwise, negative emotions are triggered when one's \added{goals} are threatened~\cite{SLazarus1991-SLAEAA}. 

In line with these theories and consistently with the operationalization adopted in our previous study~\cite{girardi2020theway}, we use emotions as a proxy for users' \textit{engagement} during product feedback interviews.  
\added{When evaluating the importance of a topic during an interview, the appraisal process of an individual is responsible for triggering an emotional reaction (i.e., engagement) based on the perceived relevance of the topic with respect to his/her goal, values, and desires.} Our choice is further supported by previous empirical findings demonstrating how emotions can be leveraged for detecting engagement in speech-based analysis of conversations~\cite{YuAW04} or to detect students' motivation~\cite{Berhenke}.  
 
Consistently with prior research on emotion awareness in software engineering~\cite{MF15,GraziotinWA15:Journal,Girardi2020}, we adopt a dimensional representation of developers' emotions.
In particular, we refer to the Circumplex Model of Affect by Russel~\cite{Russel1991}, which models emotions along two continuous dimensions, namely \textit{valence}, that is the pleasantness of the emotion stimulus, ranging from pleasant to unpleasant, and \textit{arousal}, that is the level of emotional activation, ranging from activation to deactivation.
Pleasant emotional states, such as happiness, are associated with \textit{positive} valence, while unpleasant ones, such as sadness, are associated with \textit{negative} valence.
The arousal dimension captures, instead, the level of emotional activation. Some emotions are associated with the person being inactive, thus experiencing  \textit{low} arousal, as in \textit{sadness} or \textit{relaxation}. Conversely, high levels of arousal are associated with high emotional activation, as in \textit{anger} or \textit{excitement}. 

We expect to observe different forms of engagement in relation to valence and arousal: positive-high engagement (i.e., positive valence and high arousal) may occur when users discuss topics that they consider relevant and towards which they have a positive feeling, e.g., feature users like and have an opinion they want to discuss; negative-high engagement (i.e., negative valence and high arousal) may occur when topics are relevant but more controversial, such as a feature that users do not like, or a bug they find annoying. Low engagement may occur when the user does not have a strong opinion on the topic of the discussion, and is either calm (positive valence, low arousal) or bored by the conversation (negative valence, low arousal). 
\added{In the context of the study, we aim to predict engagement through machine learning methods. Therefore, we ask participants to report their perceived engagement to build the ground truth. This means that we consider \textit{self-reported} engagement, rather than their actual engagement.}

\subsection{Biofeedback-based Classification of Emotions}
\label{sec:biofeedbackbasedclassification}
The use of physiological signals for emotion recognition has been largely investigated by affective computing research~\cite{Canento,KimA08,KoelstraMSLYEPNP12,SoleymaniAFP16,Girardi:ACII2017}.
Previous work studied the relationship between emotions and biometrics such as the electrical activity of the brain---e.g., using electroencephalogram (EEG)~\cite{Kramer90physiologicalmetrics,Reuderink:2013,SoleymaniAFP16,Li:Lu:EEG}, the electrical activity of the skin, or electrodermal activity (EDA)~\cite{Burleson:Picard, Kapoor:2007}, the electrical activity of contracting muscles measured using electromyogram (EMG)~\cite{KoelstraMSLYEPNP12,nogueira2013hybrid, Girardi:ACII2017}, and the blood volume pulse (BVP) from which heart rate (HR) and its variability (HRV) are derived~\cite{Canento,Scheirer}. 

Electroencephalogram (EEG) captures the electrical activity of the brain through electrodes placed on the surface of the scalp.
Changes in the EEG spectrum correlate with increased or decreased overall levels of arousal or  alertness~\cite{Kramer90physiologicalmetrics} as well as with the valence of the emotion experienced~\cite{Reuderink:2013, SoleymaniAFP16}.


The electrodermal activity (EDA)
measures the electrical conductance of the skin due to the sweat glands' activity.
EDA correlates with the arousal dimension~\cite{BL2008} and its variation occurs in the presence of emotional arousal and cognitive workload. Hence, EDA has been employed to detect excitement, stress, interest, attention as well as anxiety and frustration~\cite{Burleson:Picard, Kapoor:2007}. 

Heart-related metrics have been successfully employed for emotion detection~\cite{Canento,Scheirer}. In particular, blood volume pressure (BVP) measures the changes in the volume of blood in vessels, while Heart Rate (HR) and its Variability (HRV) capture the rate of heartbeats.
Significant changes in the BVP are observed in the presence of increased cognitive and mental load~\cite{kushki2011comparison}.
Increases in HR occur when the body needs a higher blood supply, for example, in the presence of mental or physical stressors~\cite{greene2016survey}.

Finally, several studies demonstrated the high predictive power of facial EMG for emotion recognition~\cite{KoelstraMSLYEPNP12,nogueira2013hybrid}. However, it leads to poor results when the sensors are placed on body parts other than the face (i.e., the arms~\cite{Girardi:ACII2017}). 

In a recent study, Girardi et al.~\cite{Girardi2020} identify a minimum set of sensors, including EDA, BVP, and HR for valence and arousal classification. 
To collect such physiological signals, they use the Empatica E4 wristband and detect developers' emotions during software development tasks. 
They found that the performance obtained using only the wristband is comparable to the one obtained using  an EEG helmet together with the wristband. 

Accordingly, in this study we use EDA, BVP, and HR collected using Empatica E4, a noninvasive device that participants can comfortably wear during interviews  (see Section~\ref{sec:materials}), thus increasing the ecological validity of our study. Furthermore, we combine biofeedback with \textit{voice} features, which were not considered in previous works.

\subsection{Voice Analysis and Classification of Emotions}
\label{sec:voiceanalysisclassification}

Classification of emotions based on the analysis of voice features is a well-developed research field, normally referred to as \textit{speech emotion recognition} (SER). Different surveys have been recently published on the topic~\cite{akccay2020speech,schuller2018speech,sailunaz2018emotion}, which highlight the maturity of research, but also point out the limits in terms of real-world applications, mainly due to limited gold standard datasets available for SER systems' training and assessment. 

Speech is composed of a diverse set of acoustic features, and its information content is usually accompanied by other so-called supporting modalities, including linguistic features (i.e., the textual content equivalent to a verbal utterance), visual features, and physiological signals such as those discussed in the previous section. 

Acoustic features used in SER are normally classified into \textit{prosodic} (e.g., pitch, tone), \textit{spectral} (i.e., frequency-based representation of the sound produced), \textit{voice quality} (e.g., measuring the stability of the voice) and Teager energy operator \textit{(TEO)-based features}, specifically developed to detect stress from the voice signal. Prosodic and spectral features are the most commonly used in the literature~\cite{akccay2020speech,sailunaz2018emotion}. In particular, the most commonly used features are Mel-scaled spectrogram and Mel-frequency cepstral coefficients (MFCCs), which are spectral features that mimic the reception pattern of sound frequency intrinsic to a human~\cite{issa2020speech}. ~\cite{issa2020speech} also uses Chromagrams---typically used for music representation---since the other features were recognised to be poor in distinguishing pitch classes and harmony~\cite{beigi2011speaker}. 

Research in SER initially focused on identifying relevant features and combination thereof to optimise the performance of traditional classification algorithms~\cite{lee2005toward,ververidis2006emotional}, leading to good recognition rates especially with Support Vector Machines~\cite{chen2012speech,schuller2004speech}. With the advent of deep neural networks, and the possibility of overcoming the feature engineering problem altogether, the focus shifts to the selection of appropriate network  architectures, and promising results are  achieved through Convolutional Neural Networks (CNNs) and Long Short-Term Memory (LSTM) models~\cite{zhao2019exploring,trigeorgis2016adieu}. To address the problem of data scarcity, the recent development of transfer learning methods has also been experimented~\cite{ottl2020group}. Another avenue of research in SER, still under development, is the combination of pure audio features with other contextual cues, including  videos~\cite{chao2015long,ren2019multi,ottl2020group}. 

Applications of the SER techniques are mostly in the field of human-computer interaction (HCI). Ramakrishnan and El Emary~\cite{ramakrishnan2013speech} list a set of ten different possible applications, including lie detection, treatment of language disorders, driving support systems~\cite{schuller2004speech}, surveillance~\cite{nanda2017neuromorphic}, and smart assistants. These latter have become commercially available in recent years---e.g., Siri, Cortana---and represent one of the natural areas of exploitation of SER research. Another traditional application,  closely related to our context, is the recognition of customer's emotions during conversations with call center operators~\cite{han2020ordinal,li2019acoustic,morrison2007ensemble,batliner2003find,petrushin1999emotion}. These works are oriented to identify critical phases of the dialogue between a customer and an artificial operator. This can be generally useful to  understand when the artificial agent is irritating the customer, thus to deciding to transfer the call to a human operator.

\subsection{Machine Learning Algorithms}
\label{sec:mlalgorithms}

\added{
To support the prediction of user engagement, we chose popular machine learning algorithms---i.e., Naive Bayes (NB), C4.5-like decision trees (DTree), Support Vector Machines (SVM), Multi-layer Perceptron for neural network (MLP), and Random Forest (RF). This choice is in line with previous studies on biometrics~\cite{FucciGNQL19,MF15,Girardi2020,vrzakova2020affect}, demonstrating the effectiveness of these algorithms for the recognition of cognitive and affective states in software engineering. In the following, we describe the main characteristics of the chosen algorithms. Detailed descriptions of the different algorithms can be found, e.g., in the book by Bonaccorso~\cite{bonaccorso2017machine}.

\paragraph{Naive Bayes (NB)} This algorithm is based on using Bayes’ theorem to calculate the probability of a data item belonging to a certain class, given its features. Bayes’ theorem is a mathematical formula that relates the conditional probability of an event to its prior probability and the evidence. NB simplifies this calculation by assuming that the input features are independent of each other, which means that the presence or absence of one feature does not affect the probability of another feature. This assumption is often unrealistic, but it makes the computation faster and easier, making it possible to handle high-dimensional and sparse data efficiently.

\paragraph{C4.5 Decision Trees (DTree)} This algorithm is based on the idea of using a tree-like structure to split the data into smaller subsets based on certain criteria. Each node of the tree represents a feature, each branch represents a possible value of that feature, and each leaf represents a class. The criteria for splitting the data are usually chosen to maximize the information gain, which measures how much each split reduces the uncertainty or impurity of the data. DTrees are easy to interpret and can handle both categorical and numerical features, but they are prone to overfitting, which means that they may memorize the training data and fail to generalize to new data. 

\paragraph{Support Vector Machines (SVM)} SVMs are based on the idea of finding the optimal hyperplane that separates the data into two classes with the maximum margin. A hyperplane is a subspace of one dimension less than the original space, and a margin is the distance between the hyperplane and the closest data points from each class. For example, if we want to classify two-dimensional data into two classes, we can find a line that divides them such that there is as much space as possible between the line and the nearest points from each class. SVMs can also handle non-linear data by using kernel functions that map the data into a higher-dimensional space where they become linearly separable. SVMs typically achieve high accuracy and robustness with relatively few parameters.

\paragraph{Multi-layer Perceptron (MLP)} This is a type of artificial neural network, and is based on the idea of using multiple layers of nodes connected by weighted edges to learn complex non-linear patterns from the data. Each node applies an activation function to the weighted sum of its inputs and passes the output to the next layer. The activation function determines how much each node contributes to the output, and it can be linear or non-linear. MLP can learn from data by adjusting the weights through backpropagation, which is an algorithm that propagates the error from the output layer back to the input layer and updates the weights accordingly. Given the layer-based structure, MLP can learn complex abstractions, each one incrementally ``embedded'' in one of the layers during the training phase. 

\paragraph{Random Forest (RF)} This algorithm combines multiple DTrees and outputs the class that is the mode of the individual trees. A RF is an ensemble classifier that reduces the variance and overfitting of a single DTree by using random subsets of features and data for each tree. For example, if we want to classify items based on their features, we can create multiple decision trees that use different combinations of features, and different subsets of data, for each tree. Then, we can take a majority vote among all the trees to decide which class to assign to a new item. RF can handle large datasets with high dimensionality and missing values, because it can average out the noise and errors from individual DTrees.

} 

\subsection{Related Work}
\label{sec:biofeedbackvoiceinSE}
Biometric sensors have been leveraged in several software engineering studies for recognizing the cognitive and affective states of software developers. 

In one of the early studies in this field, Parnin~\cite{Parnin2011} envisions an approach to infer developers' cognitive states based on the analysis of sub-vocal signals, that is the electrical signal the brain transfers to the mouth and vocal cords while performing complex cognitive activities. While presenting preliminary findings, this study demonstrates that it is possible to use EMG to capture the subvocalization associated with programming and leverage this information to distinguish between easy and hard development tasks. 
Fucci et al.~\cite{FucciGNQL19} use multiple biometrics, including EEG, EDA, HR, and BVP, to distinguish between code and prose comprehension tasks during software development. 
Fritz et al.~\cite{FritzBMYZ14} employ EEG, BVP, and eye tracker to measure the difficulty of programming tasks, thus preventing developers from introducing bugs. 
The authors use the same set of sensors in a follow-up work aimed at classifying the emotional valence of developers during programming tasks~\cite{MF15}. 
Girardi et al.~\cite{Girardi2020} replicate previous findings by M{\"u}ller and Fritz~\cite{MF15} regarding the use of non-invasive sensors for valence classification during software development tasks.
Furthermore, Girardi et al. also address the classification of arousal. In a recently published work, Girardi et al.~\cite{girardi2021emotions} study the relationship between biometric signals and productivity of developers with an \textit{in vivo} setting involving engineers at work. 
Other studies also propose approaches for predicting developers' interruptibility~\cite{ZugerMMF18} and for identifying code quality concerns~\cite{MullerF16} by leveraging EDA, HR, and HRV.

Biofeedback has also been used in RE, mainly to capture users' emotions \textit{while} using an app.
For example, some studies~\cite{SMK19,MSE19} use mobile phone cameras 
to recognize facial muscle movements and associate them with the users' emotions when using different features of an app. 
This methodology was recently applied to enable user validation of new requirements~\cite{SKE19} and to identify usability issues~\cite{JBB19} with minimal privacy concerns~\cite{SSM19}. Part of the authors of the current paper previously proposed using biometrics in requirements elicitation interviews~\cite{spoletini2016empowering}. Our previous work focused on ambiguity, and remained at the research preview stage, as it evolved in the current work after pilot experiments. Other work, in affective RE, \added{acknowledges} the primary relevance of users' emotions~\cite{Sut11}, and generally focuses on the application of sentiment analysis techniques to textual user feedback, for example from tweets or app reviews~\cite{GAS17,MM19}. Other uses of sentiment analysis in RE include the prediction of ticket escalation in customer support systems~\cite{WLD19}. Emotions have also been considered in early-stage RE activities, such as requirements elicitation and modelling. For example, Colomo-Palacios et al.~\cite{CCS11} asked users to rank requirements according to Russel's Valence-Arousal theory, which is the one that we adopt in the present study. Other researchers leverage information regarding users' emotions gathered through psychometrics (e.g., surveys) to augment traditional requirements goal modelling approaches~\cite{TSP19,MPL15} and artefacts, such as user stories~\cite{KS17}.


\paragraph{Contribution}
Compared to previous works using biofeedback and voice analysis in software engineering and RE, 
our study is among the first ones to specifically focus on product feedback interviews rather than product usage or development tasks.
Previous studies (e.g.,~\cite{SMK19,MSE19}) focus on detecting the user's engagement experienced \textit{while} using the software features. 
In our case, we aim to detect users' engagement about certain features when users reflect on the features and speak about them. This captures a different moment---a verbalized, more rational one---of the relationship between the user and the product. 
Furthermore, in interviews we can consider \textit{what if} scenarios (e.g., financial and privacy-related questions in Table~\ref{tab:interviewquestions}), which is not possible when performing observations without interacting with users.
Finally, to our knowledge, our work is among the first ones that use voice features to predict the emotion of a speaker involved in a software engineering activity.

%% file: section/design.tex
\section{Research Design}
\label{sec:design}


Our study is \textit{exploratory} in nature, aiming to investigate a certain area of interest---i.e., \textit{engagement} in product feedback interviews---and identify possible  avenues of research. We adopt a quantitative experimental approach involving human subjects, and oriented to compare software-based artifacts (i.e., machine learning algorithms and feature configurations). The study was approved by the Kennesaw State University review board (study 16-068).


\label{sec:questions}
The main goal of this study is to understand to what extent we can use biofeedback devices and voice analysis to predict users' engagement during interviews. \rev{We focus on interviews in which users provide feedback about product features. This type of interview is assumed to be more structured than initial requirements elicitation interviews oriented to product design, and thus more amenable to be treated in quantitative experimental design. We interview users in controlled settings about the features of a representative software, namely Facebook. Although our study is not primarily oriented to consumers' products, we selected Facebook as discussion topic based on the following assumptions: 1) the platform is widely known and used, so participant selection is simplified; 2) the platform embeds several functionalities that can be found in other products (messaging, information sharing, networking, etc.). We design a structured interview script specifically oriented to provoke mild emotions, as the ones that one could expect in an interview context. Though interviews can in principle be unstructured, a fixed script facilitates data analysis, as data from different subjects can be isolated more easily and treated homogeneously.
The scripts are designed to discuss topics that can typically emerge in interviews with users.}
\rev{Moreover---on top of being a widespread application---at the time of performing the study Facebook was still considered an uncontroversial platform.}



The research questions (RQs) of the study are the following:
\begin{itemize}
    
    

    \item \textbf{RQ1:} \textit{To what extent can we predict users' engagement using biofeedback measurements and supervised classifiers?}  
    With this question, we aim to understand whether it is possible to automatically recognize engagement with biofeedback. More specifically, we want to assess to what extent we can recognize emotional valence and arousal---i.e., the two dimensions we use for the operationalization of engagement. To collect trainig and testing data, we first interview Facebook users, asking their opinion about the platform. After the interview, we ask them to report their engagement for each of the different questions. During the interviews with users, we acquire their raw biofeedback signals. We use features extracted from the signals, and consider intervals of reported engagement as classes to be predicted. Based on these data, we evaluate and compare different supervised machine learning classifiers.
    
    \item \textbf{RQ2:} \textit{To what extent can we predict users' engagement using voice analysis and supervised classifiers?} This question aims to understand whether we can recognize engagement with automatic voice analysis.  This RQ is motivated by the need to support a different use case in which biofeedback devices might not be used. It is the case, for example, of interviewees being uncomfortable wearing a tracking device during the conversation. In other situations, the use of physical devices might even be unfeasible, due to remote interview settings. 
    To this extent, we record the audio of the interviews, and we extract voice features from the audio signals. We then use the voice features to train and compare the previously used supervised classifiers. 
    
    \item \textbf{RQ3:} \textit{To what extent can we predict users' engagement by combining voice and biofeedback features?}
    This question aims to investigate the use of voice and biofeedback features in conjunction. By training and comparing the classifiers, we check if and in which way the combination of features improves  the performance of the approach.

\end{itemize}

\subsection{Study Participants} 
\label{sec:studycontext}
We recruited 31 participants among the students of Kennesaw State University with an opportunistic sampling. 
The participation was not restricted by major or academic level, but the only main requirement was to be an active Facebook user (access to Facebook at least once per day, self-declared), as the user interview questions dealt with this social network. 
More than 90\% of the participants were undergraduate students divided in 11 majors.
To account for differences in biometrics due to physiological aspects~\cite{bent2020investigating}, we try to have a pool of participants as varied as possible by including multiple ethnic groups and both female and male subjects. Specifically, approximately 65\% of the participants were male, and their ages varied between 18 and 34 with both median and average equal to 22.
Participants were either native speakers or proficient in English. The majority (58\%) were white/Caucasian, 23\% black/African American, 13\% Hispanic/Latino, and the remaining 6\% was Asian/Pacific islander.
During the data analysis, we removed 10 participants because either the collected data were incomplete or the available information were not considered reliable (e.g., they provided the same response to all the questions in the surveys).
Of the remaining 21 participants, approximately 67\% were male with the following racial/ethnicity distribution, 67\% white/Caucasian, 14\% black/African American, 14\% Hispanic/Latino, and 5\% Asian/Pacific islander. 
We collected information about the ethnicity of participants because the research demonstrated that heart-rate optical sensors might give more/less reliable readings based on the skin tones. Having a diverse pool of participants in terms of ethnicity strengthens the validity of our empirical findings.
Participants received a monetary incentive of \$25 for up to one hour of their time.

\subsection{Biofeedback Device and Signals} 
\label{sec:materials}
The device we use to acquire the biofeedback is the Empatica E4\footnote{\url{https://www.empatica.com/research/e4/}} wristband.
We selected it as it is used in several studies in affective computing~\cite{greene2016survey} as well as in the field of software engineering~\cite{MF15,FucciGNQL19}.
Using the Empatica E4, we collected the following signals:
\begin{itemize}
    \item \textbf{Electrodermal Activity}: 
    EDA can be evaluated based on measures of skin resistance. Empatica E4 achieves this by passing a small amount of current between two electrodes in contact with the skin, and measuring electrical conductance (inverse of resistance) across the skin. EDA is considered a biomarker of individual characteristics of emotional responsiveness and, in particular, it tends to vary based on attentive, affective, and emotional processes~\cite{Critchley2013}. 
    \item \textbf{Blood Volume Pulse}: BVP is measured by Empatica E4 through a photoplethysmography (PPG)---an optical sensor that senses changes in light absorption density of the skin and tissue when illuminated with green and red lights~\cite{allen2007photoplethysmography,sinex1999pulse}. 
    \item \textbf{Heart Rate}: HR is measured by Empatica E4 based on the elaboration of the BVP signal with a proprietary algorithm.
\end{itemize}

Research identified a minimal set of biometrics for reliable valence and arousal detection, consisting in the EDA, BVP, and HR measured by the E4 wristband~\cite{Girardi2020}.  

\subsection{Audio Device and Signals} 
\label{sec:audiorecording}

The interviews' audios were captured using the default audio recorder of a Mac OS laptop, and the files were stored in the classical Waveform Audio File Format (i.e., {\tt .wav}), which is an uncompressed representation of the raw signal. Audio is a complex, information-rich signal, and a largely variable set of classical features are used to characterise its salient aspects~\cite{shuller2013}. Among these features, we consider the following ones:  

\begin{itemize}
    \item \textbf{Mel Spectrogram}: the acoustic time-frequency representation of sound.  
    \item \textbf{Mel-Frequency Cepstral Coefficients (MFCCs)}:  
    MFCCs are the  representation of the short-term power spectrum of sound. More in details, Cepstral features contain information about the rate changes in the different spectrum bands and they have the ability to separate the impact of the vocal cords and the vocal tract in a signal. In the MFCC, these features are extracted at a frequency more audible by human ears.
    \item \textbf{Chromagram}: the Chromagram is a 12-element feature vector indicating how much energy of each pitch class is presented in the signal. This is typically used to model harmonic and melodic characteristics of music, and it is recognised as useful also to model the emotional aspect of voice~\cite{shuller2013}.
\end{itemize}    

We choose these features as they are amongst the most common in speech emotion recognition~\cite{issa2020speech,shuller2013}.





\subsection{Experimental Protocol and Data Collection}
\label{sec:protocol}
Three main roles are involved in the experiment: \textit{interviewer}, \textit{user}, and \textit{observer}.
The interviewer leads the experiment by asking questions to the user, while the observer tracks the interview by annotating timestamps of each question, monitoring the output of the wristband, checking that the audio recording is operational, and annotating general observations on the interview and behaviour of the user. 

The experimental protocol consists of four phases: (i) device calibration and emotion triggering, (ii) user's interview, (iii) self-assessment questionnaire, and (iv) wrap-up. At the beginning of the experiment, we explained the different steps to the participants, who had the opportunity to ask clarification questions throughout the experiment. \added{Specifically, we introduced the experiment verbally along the following lines: ``Elicitation is a phase of the software engineering process and aims at discovering user needs for a new or modified software product. In the literature, there exists a variety of elicitation techniques. Among those, unstructured interviews are the most commonly used. An interview involves two roles: a user and a requirements analyst. In this research study, you will play the user's role and be asked by an analyst to provide your opinions about Facebook. Imagine that your feedback will be used to improve the platform. The experiment is composed of different parts. In the first part, you will be asked to watch a slideshow composed of 35 images. Then, you will complete a simple demographic survey and a survey on the images you just watched. The third step consists of an interview about Facebook, and, finally, you will be asked to fill out a survey about the interview, and your emotional involvement with the topics discussed. You will be asked to wear a wristband to measure your heart rate, temperature, and skin galvanic response, and the interview will be recorded.''} 

\added{In this phase, the participants were also provided with the concepts related to engagement, and how to recognise it. To simplify the concept for the participants, we indicated engagement with the term ``emotional involvement''. Definitions were given to explain the concepts of ``degrees of (emotional) involvement'' (arousal) and ``quality of (emotional) involvement'' (valence). The definitions were given verbally along the following lines: ``You have a strong involvement when you feel that a certain topic is particularly relevant to you, and triggers an emotional reaction. This reaction can be positive, such as enthusiasm, or negative, such as anger or disappointment, and this indicates the quality of your involvement. You have low involvement when you do not have a strong emotional reaction. In this case, the quality of your involvement can be positive, which means that you feel quiet, or negative, which means that you feel bored''.} According to the IRB that approved the experiment protocol, the collected data does not present more than minimal risks to the participants.

\subsubsection{Device calibration and emotion triggering}
\label{sec:calibrationtask}
In line with previous research~\cite{MF15,Girardi2020} we run a preliminary step for device calibration and emotion elicitation.
The purpose of this phase is threefold. 
First, we want to check the correct acquisition of the biofeedback signal by letting the wristband record the raw signals for all sensors under the experimenter's scrutiny. 
Second, the collected data will be needed to adjust the scores obtained during the self-assessment questionnaire (see Sect.~\ref{sec:datacleaning}).
Third, we want the participants to get acquainted with the emotion self-report task.

Accordingly, we run a short emotion elicitation task using a set of emotion-triggering pictures.
Each participant watches a slideshow of 35 pictures. 
Each picture is displayed for 10 seconds, with intervals of five seconds between them to allow the user to relax.
The whole slideshow lasts for nine minutes.
During the first and last three minutes, calming pictures are shown to induce a neutral emotional state, while during the central 3 minutes the user sees pictures aimed at triggering negative and positive emotions. 
The pictures have been selected from the Geneva database~\cite{dan2011geneva}, previously used in software engineering studies~\cite{MF15}. 
The user is then asked to fill out a form to report the degree of arousal and valence they associated with the pictures on a visual scale from 0 to 100. 
As done in previous work~\cite{MF15}, for each picture, the user is asked \added{to assess two items}: \textit{``You are judging this image as: ''} (0 = Very Negative; 50 = Neutral; 100 = Very Positive); \textit{``Confronted with this image you are feeling: ''} (0 = Relaxed, 50 = Neutral, 100 = Stimulated). 


\input{tables/interview-table}

\subsubsection{User's Interview}
\label{sec:interviewtask}
A trained interviewer conducted the interview with each participant. 
The interview script consists of 38 questions concerning the Facebook platform. 
Questions are grouped into seven topics---i.e., \textit{usage habits}, \textit{privacy}, \textit{procedures}, \textit{relationships}, \textit{information}, \textit{money}, and \textit{ethics}.
The questions are reported in Table~\ref{tab:interviewquestions}. For each topic, we include  multiple questions, to allow users sufficient time to get immersed in the topic, and collect more stable biometric parameters in relation to the topic. 
Questions related to topics we expect to raise more engagement, (i.e., privacy, relationship, money, and ethics) are separated by questions on topics that are expected to reduce user engagement (i.e., usage habits, procedures, and information).
The lower degree of engagement for the latter topics was assessed during preliminary experiments in which the questions were drafted and finalised\footnote{During the experiments reported in this paper, we saw that \textit{usage habits} was associated with higher engagement, instead. Discussion on this aspect is reported in Sect.~\ref{sec:discussion}.}. Overall, our instrument has been designed to provoke mild emotions, triggered by reflections about the product features, so to measure if these product-related emotions could be captured by biometric data. We have chosen to use a structured interview questionnaire to have comparable interviews, so to facilitate subsequent analysis. To improve the degree of realism of the interview, we did not constrain the time slot for answering each question, so that the interviewees could freely express their opinions.  

During the interview, the wristband records the biofeedback parameters, the audio recorder acquires the voice of the speakers, and the observer annotates the timestamp of each question. 
We use this information to align the sensor data with the questions.
Based on a preliminary run, each interview was estimated to last for about 20 minutes.

\subsubsection{Self-assessment Questionnaire and Wrap-up}
\label{sec:interviewquestionnaire}
For each question in the interview script (i.e., $Q_i$), the interviewer asks the participant to report their involvement using two 10-point rating scale items: $q_A(Q_i)$: How much did you feel involved with this topic? (1 = Not at all involved; 10 = Extremely involved); $q_V(Q_i)$: How would you rate the quality of your involvement? (1 = Extremely negative; 10 = Extremely positive). 
These two questions aim at measuring the engagement of the user in terms \textit{arousal} ($q_A$) and \textit{valence} ($q_V$). 
Afterwards, the observer downloads and stores the wristband data as well as the voice recording and the  questionnaires filled by the participant. 
The wristband memory is then erased to allow further recording sessions. 

\subsection{Pre-processing, Feature Extraction, Classification}
\label{sec:datacleaning}
The data from the interview questionnaire are used to produce the gold standard---i.e., the labels for valence and arousal to be predicted. 

We define \textit{positive}, \textit{negative}, and \textit{neutral} labels for valence, and \textit{high}, \textit{low}, and \textit{neutral} labels for arousal. \added{The reason for transforming the scores of the Likert scales used by the participants into three classes for valence and arousal is twofold. On the one hand, we considered that, from the participant's point of view, it would have been a too strict constraint to express their emotions only in terms of positive/negative/neutral valence and low/medium/high arousal. On the other hand, we considered that predicting specific values of the Likert scale would not have been feasible with the limited number of data points collected. Therefore, we opted for the compromise of asking participants to express their engagement on a Likert scale, and then we discretised the values into three classes for each dimension.}

We discretise the scores in the Likert rating scale following an approach utilized in previous research~\cite{MF15,Girardi2020}. 
First, we adjusted the valence and arousal scores based on the mean values reported while participants watched emotion-triggering pictures (see Section~\ref{sec:calibrationtask}).
This step is necessary to take into account fluctuations due to individual differences in the interpretation of the scales in the interview questionnaire.
Then, we perform a discretization of the values into the three categories (i.e., labels) for each dimension using k-means clustering\footnote{We use the k-means implementation in by the \texttt{arules} R package.}. 


To synchronize the measurement of biofeedback and voice signals with the self-assessment, we (1) save the timestamp corresponding to the interviewer asking question $Q_i$  (i.e., $timestamp(Q_i)$), 
(2) calculate the timestamp associated to the next question $Q_{i+1}$ ($timestamp(Q_{i+1})$), and 
(3) select each signal samples recorded between $timestamp(Q_i)$ and $timestamp(Q_{i+1})$.

For each interview question $Q_i$, we have: 
\begin{itemize}
    \item a set of biofeedback signal samples (for EDA, BVP, and HR) within the time interval associated with $Q_i$; 
    \item a voice signal sample in the form of a {\tt .wav} file---the segment of the {\tt .wav} file of the whole interview for the time interval associated to $Q_i$;
    \item two labels, one representing arousal ($q_A(Q_i)$) and the other representing valence ($q_V(Q_i)$) according to the self-assessment questionnaire.
\end{itemize}

The labels are used to form the gold standard to be predicted by the algorithms based on features extracted from the signal samples.


To maximize the signal information and reduce noise caused by movements, we apply multiple filtering techniques.
Regarding BVP, we extract frequency bands using a band-pass filter algorithm  at different intervals~\cite{Canento}. 
The EDA signal consists of a tonic component (i.e., the level of electrical conductivity of the skin) and a phasic one representing phasic changes in electrical conductivity or skin conductance response, or SCR~\cite{BWJR15}. 
We extract the two components using the cvxEDA algorithm~\cite{GVLSC16}. 

\begin{table}[]
\caption{Machine learning features grouped by physiological and voice signal.}
\label{tab:features}
\centering
\begin{tabular}{p{0.6cm}p{8.5cm}}
\textbf{Signal} & \textbf{~~Features} \\
\hline
EDA & \begin{tabular}[c]{@{}l@{}}- mean tonic\\ - phasic AUC \\ - phasic min, max, mean, sum peaks amplitudes\end{tabular} \\
\cline{2-2} 
BVP & \begin{tabular}[c]{@{}l@{}}-  min, max, sum peaks amplitudes\\ - mean peak amplitude (diff. between baseline and task)\end{tabular} \\
\cline{2-2}
HR & \begin{tabular}[c]{@{}l@{}}- mean, sd. deviation (diff. between baseline and task) \end{tabular} \\
\cline{2-2}
Voice & \begin{tabular}[c]{@{}l@{}}- mean Mel Spectrogram \\ - mean MFCC (mean of the first 20 MFCC features) \\ - mean Chromagram (mean of the 12 Chroma features) \end{tabular} \\
\hline
\end{tabular}
\vspace{-5mm}
\end{table}

After signals pre-processing, we extracted the features presented in Table~\ref{tab:features}, which we use to train our classifiers. 
We select biofeedback features based on previous studies using the same signals~\cite{MF15,FucciGNQL19,Girardi2020} and we choose audio features according to recommendations from the specialised literature~\cite{shuller2013,issa2020speech}.
For the audio signal, we use the Python package Librosa~\cite{mcfee2015librosa}\footnote{https://librosa.org}.

We address the problem of predicting user engagement using machine learning classifiers.  
\added{In line with previous research~\cite{FucciGNQL19,MF15,Girardi2020,vrzakova2020affect}, we chose popular algorithms---i.e., Naive Bayes (NB), C4.5-like decision trees (DTree), Support Vector Machines (SVM), Multi-layer Perceptron for neural network (MLP), and Random Forest (RF).}  

\subsection{Analysis Procedure} 
\label{sec:analysis}
The analysis procedure aims to answer the three RQs. For each user, we use the biometrics gathered in the interview phase as input features for the different classifiers. We first consider solely biofeedback features (RQ1), then voice features (RQ2) and finally their combination (RQ3). 


In line with previous research~\cite{MF15,Girardi2020}, we target a \textit{binary} classification task.
In particular, we distinguish between \textit{positive} vs \textit{negative} valence, and \textit{high} vs \textit{low} arousal. As such, we exclude the \textit{neutral} label from the gold standard and focus on more polarised values. Although this reduces our dataset, it also facilitates the separation between clearly distinguished emotional states\footnote{Preliminary experiments were performed considering a 3-label problem, but the number of vectors resulted too small to achieve acceptable results. This aspect is further discussed in Sect.~\ref{sec:threats}.}.

We evaluate our classifiers in the \textit{hold-out} setting. Therefore, we split the gold standard into train (70\%) and test (30\%) sets using a stratified sampling strategy, which allows having a balanced set of instances from the different classes in both sets.  
For each algorithm, we search for the optimal hyper-parameters~\cite{TMH16, TMH18} using leave-one-out cross-validation on the train set---i.e., the recommended approach for small training sets~\cite{Raschka} such as ours.
The resulting model is then evaluated on the test set to assess its performance on unseen data and avoid overfitting.
We repeat this process 10 times with different splits of the train and test sets to further increase the validity of the results. The performance is then evaluated by computing the mean for precision, recall, F1-measure, and accuracy over the different runs.
This setting is directly comparable to the one implemented by previous work~\cite{MF15,Girardi2020}, which includes data from the same subject in both training and test sets. 

The process outlined above is repeated with a maximum of eight different settings, based on the three following data pre-processing steps, aimed at improving the performance of the machine learning algorithms without losing the validity of the results:

\begin{itemize}
    \item \textbf{Standard Scaling:} the features in the training set are standardised so that their distribution will have a mean value of 0 and standard deviation of 1. The standardisation parameters from the training set are then applied to scale the test set. This way, information from the test set (i.e., its standard deviation) is not passed to the training set, which could bias the learning process. Standard scaling is essential for machine learning algorithms that calculate distances between data, in our case SVM and MLP. If not scaled, the feature with a higher value range starts dominating when calculating distances. Scaling should not affect rule-based algorithms that consider each feature separately, and are not affected by monotonic transformations of the variables, such as standard scaling. Standard scaling is performed using the \texttt{StandardScaler} available in \texttt{scikit-learn}.
    
    \item \textbf{Balancing:} Synthetic Minority Oversampling Technique (SMOTE) is a traditional data augmentation technique applied to train machine learning models in case of class imbalance~\cite{chawla2002smote}. In case of class imbalance, machine learning algorithms tend to perform poorly on the minority class, as they do not have a sufficient amount of examples  to learn from and build a fair classification model. To overcome this issue, SMOTE creates synthetic examples of the minority class, in our case based on the k-nearest neighbour algorithm. To prevent data leakage, SMOTE is applied solely to the training set, therefore the test set does not contain synthetic data. Balancing is performed using the SMOTE implementation in the \texttt{imblearn} Python package. 
    
    \item \textbf{Imputation:} data imputation is normally adopted when some features have missing data. In our case, we miss voice feature data for 80 arousal vectors and 45 valence vectors. However, we can infer (\textit{impute}) the data by using the corresponding biofeedback features. Imputation is performed using k-nearest neighbors approach implemented in the \texttt{KNNImputer} from \texttt{scikit-learn}. 
\end{itemize}

%% file: tables/interview-table.tex
\begin{table*}[]
\centering
\caption{List of questions asked during the Interview Phase}
\label{tab:interviewquestions}
\begin{scriptsize}
\begin{tabular}{|p{1\textwidth}|}

\hline

\textbf{USAGE HABITS}                                    \\ \hline
1. Do you use the Facebook chat function? \\
2. (If yes to 1) Who are the people you talk to most frequently using the Facebook chat? (If no to 1) Do you use any other chat applications? \\
3. How many hours do you use Facebook per day? \\
4. When you check Facebook, what is the average length of time you spend per session? \\
5. Is Facebook your primary source of social media? (If yes, why? If no, what other social media you use more often? Why is it superior?) \\ \hline

\textbf{PRIVACY}                  \\ \hline
6. If someone shared a photo of you in an embarrassing, incriminating, or shameful situation, how would you react? (Do you think Facebook has a responsibility to prevent it from happening? Should they be allowed to remove the photo on your behalf?) \\
7. If someone tagged you in a post which contained topics you are not comfortable sharing on Facebook (e.g., your political view, sexual preference, …), how would you react? (Do you think Facebook has a responsibility to prevent it from happening?) \\
8. How would you feel knowing that someone (e.g. your SO) accessed your profile and searched it? \\
9. Imagine Facebook begins using profile information to generate ad content. Would you be okay with this? (why?) \\
10. In relation to Facebook, what is private information?  \\ \hline

\textbf{PROCEDURE} \\ \hline
11. Can you explain me how to add a new friend on Facebook? \\
12. Can you explain me how to find Facebook pages that match your interest?\\
13. Can you explain to me how to block a person on Facebook?  \\ \hline

\textbf{RELATIONSHIP}  \\ \hline
14. Are you connected on Facebook with members of your family? (If so, do you interact with them using Facebook? If not, why?)\\
15. Have you ever had a family member (even of your extended family) delete you from his/her friend list? (If so why?)\\
16. Have you ever wanted to delete or deleted a family member (even of the extended family) from your set of friends? (If so why?)\\
17. Have you ever used Facebook to begin a long-distance relationship with someone you could not realistically meet? (If so, tell us about it.)\\
18. Have you ever considered ending a friendship/relationship over their Facebook behavior? (What did they do to make you consider this?) \\ \hline

\textbf{USAGE HABITS}  \\ \hline
19. Do you use Facebook using the mobile app or your PC?\\
20. Do you post regularly on the dashboard?\\
21. Do you click on posts that link to other websites? \\ \hline

\textbf{PROCEDURE} \\ \hline
22. Can you explain to me how to set the privacy settings?\\
23. Can you explain to me how to change the password? \\ \hline

\textbf{MONEY} \\ \hline
24. Would you agree to pay a subscription to use Facebook? If yes, how much would you consider a reasonable amount to pay? (If not, why?)\\
25. If the application for PC available from your browser was free, but the mobile app was not. Would you pay for it?\\
26. Suppose that the free access to Facebook was limited in time, information you can access or which version of the app you can use. Which of these functionalities would have to be excluded from the free version for you to be interested in the subscription? Why that Specific one?\\
27. If Facebook would pay you in exchange for you performing tasks or taking surveys, would you be interested in them? (If yes, for how much? If the tasks could be considered unethical, would you still do it?) \\
28. Suppose Facebook will become a subscription service starting from tomorrow and you decide not to pay. What should Facebook do with your profile and data? \\
\hline

\textbf{INFORMATION}  \\ \hline
29. When you read something that you find interesting, do you share it?(What motivates you to share it? Are you likely to share something without reading it?)\\
30. Is the information on Facebook more or less reliable than other sources? (For what reason?)\\
31. What is inappropriate information for Facebook? (Is there any information that should never reach Facebook? Should Facebook be used as a news source?) \\ \hline

\textbf{PROCEDURE} \\ \hline
32. Can you explain to me how create a post and tag someone into it?\\
33. Can you explain to me how to find friends that have no mutual friends? \\ \hline

\textbf{ETHICS} \\ \hline
34. FB censures some photos and posts if their content is signaled as inappropriate. Do you think this is correct? Where should the line be drawn between censure and freedom?\\
35. Recently FB has censored pictures of women breastfeeding even if the breast was not visible? Why do you think they do this? Should they be allowed to?\\
36. Recently FB workers admitted to routinely suppressing conservative content, do you feel they did anything wrong? (Why or why not?)\\
37. Should FB play a role in limiting/removing hate speech from the site? Is it ethical if they do?\\
38. Terrorist groups are known to have very active social media presences. Suppose Facebook began submitting information from all profiles to the government for help in tracking these groups. Would you be okay with this? Why? \\ \hline

\end{tabular}
\end{scriptsize}

\end{table*}

%% file: section/results.tex
\section{Execution and Results}
\label{sec:results}

\begin{figure}[h]
\centering
\includegraphics[width=0.7\columnwidth]{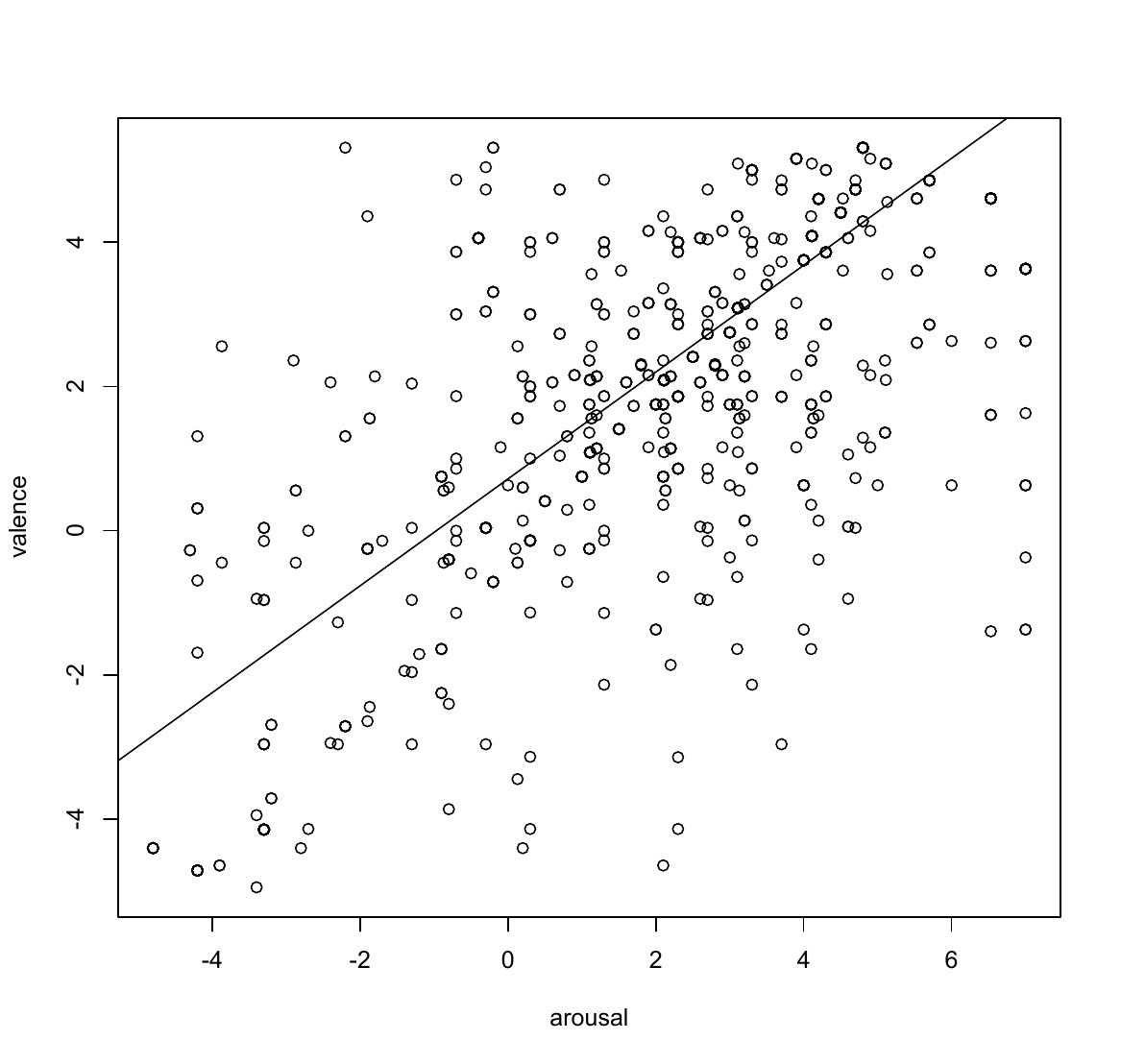}
\caption{Scatter plots of normalised valence and arousal according to the self-assessment questionnaire.}
\label{fig:valencearousalscatter}
\centering
\end{figure}

The data were initially gathered from 31 participants. Interviews lasted 18 minutes on average \added{(excluding the 9-minute calibration phase)}. We discarded the data from those subjects for which data were largely incomplete \added{(i.e., the recording of the wristband data was only partial due to recording errors, 4 subjects)}, or that appeared to have a low degree of standard deviation (i.e., lower than one) in valence and arousal (\added{4 subjects}). Although these subjects may, in principle, have had little variations in their actual emotions, they can be considered outliers with respect to the rest of the subjects. As data are treated in aggregate form, and given the limited number of data points, including these outliers could have introduced undesired noise. We also discarded data whenever some inconsistency was observed through the different pre-processing steps. For example, our protocol required manual annotations of the timestamps during the interview, which sometimes led to no plausible timestamps. In these cases (\added{2 subjects}), we had to remove the data. At the end of this process, we produced the feature vectors and associated labels for valence and arousal (776 vectors in total from \added{21 subjects}). The scatter plot for the two dimensions is reported in Fig.~\ref{fig:valencearousalscatter}. 
The normalised range of the labels, evaluated by means of k-means clustering as explained in Sect.~\ref{sec:analysis}, is as follows. For valence we have: [-4.94,-1.03) \textit{negative};  [-1.03,2.52) \textit{neutral};  [2.52,5.31] \textit{positive}. For arousal we have:  [-4.8,0.308) \textit{low}; [0.308,3.57) \textit{neutral}; [3.57,7] \textit{high}. 

\begin{table}[h!]
\caption{Label distribution in the gold standard for biofeedback feature vectors and for experiments using imputation.}
\label{tab:goldstandard}
\centering
\begin{tabular}{|c|c|c|c|c|c|}
\hline
\multicolumn{3}{|c|}{\textit{Arousal}} & \multicolumn{3}{c|}{\textit{Valence}}           \\ \hline \hline
\textbf{High} & \textbf{Low} & Neut. & \textbf{Positive} & \textbf{Negative} & Neut. \\ \hline
245  (66\%)     & 191 (44\%)      & 340  & 345  (79\%)          & 89  (21\%)          & 342   \\ \hline
\end{tabular}

\end{table}

\begin{table}[h!]
\centering
\caption{\added{Label distribution in the gold standard for voice feature vectors and combined feature vectors without imputation.}}
\label{tab:goldstandardaudio}
\begin{tabular}{|c|c|c|c|c|c|}
\hline
\multicolumn{3}{|c|}{\textit{Arousal}} & \multicolumn{3}{c|}{\textit{Valence}}           \\ \hline \hline
\textbf{High} & \textbf{Low} & Neut. & \textbf{Positive} & \textbf{Negative} & Neut. \\ \hline

  215 (61\%)     & 141 (39\%)      & 303  & 316 (81\%)          &  73  (19\%)          & 270  \\ \hline
\end{tabular}

\end{table}

As our goal is to discriminate between high (positive) and low (negative) arousal (valence), we removed all the items for which the label resulted \textit{neutral} for the dimensions, based on the participant's answers. Therefore, our gold standard includes only the vectors labelled as high (positive) or low (negative) and we model our problem as a binary classification task. Table~\ref{tab:goldstandard} reports the gold standard dataset with valence and arousal distribution, when considering biofeedback features (for RQ1). \added{After the removal of neutral cases, the remaining data still referred to 21 subjects.}
Voice feature vectors corresponding to each biofeedback vector could not be identified for part of the gold standard items, as the audio recording was unreliable for some subjects (\added{2 subjects}). Therefore, the gold standard dataset for audio only (RQ2) and for combined features (RQ3) without imputation is a subset of the original gold standard, and is reported in Table~\ref{tab:goldstandardaudio} (\added{19 subjects in total, 659 vectors). After removing neutral cases for voice, the remaining data still referred to 19 subjects}.

\begin{figure}[h]
\includegraphics[width=0.8\columnwidth]{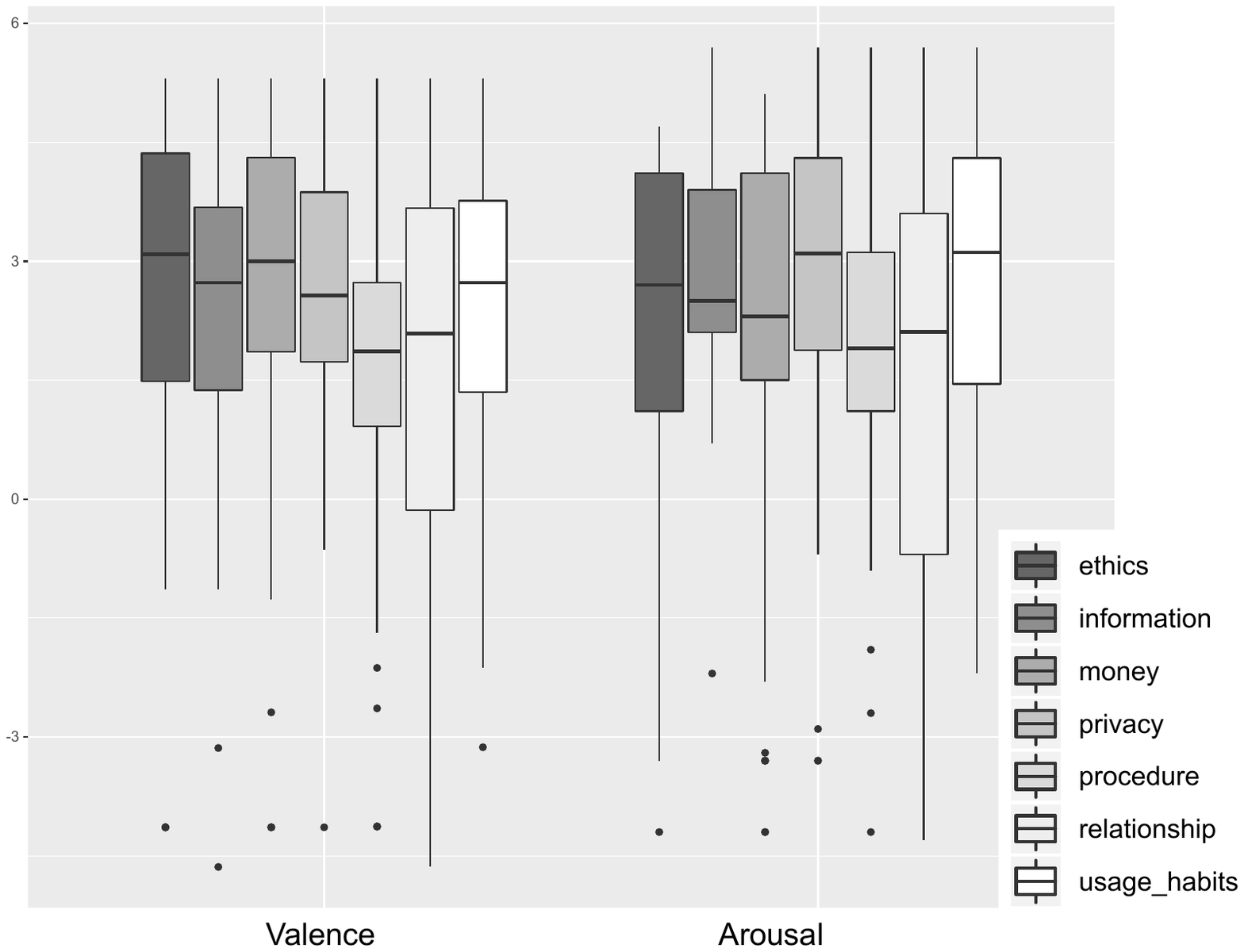}
\caption{Box plots of normalised valence and arousal for each group of questions, according to the self-assessment questionnaire.}
\label{fig:valencearousalbyquestion}
\centering
\end{figure}

\subsection{Descriptive Statistics}
\label{sec:descriptivestatistics}

In the following, we report descriptive statistics on the data. 
Table~\ref{tab:statisticsreports} reports the ranges of valence and arousal, according to the self-assessment questionnaire. We report both original values and normalised ones (``norm'', in the table). 
We see that, overall, users tend to give high scores both for arousal and valence (both averages are above 7), indicating that the interview is generally perceived as \textit{positively engaging}. Although they used the whole 1 to 10 scale for both dimensions, indicating that the interview appeared to cover the whole range of emotions, we see that the standard deviation is not particularly large, especially for valence. Indeed, considering the 1 to 10 scale, the value of standard deviation (Std. Dev. in Table~\ref{tab:statisticsreports}) indicates that around 68\% of the subjects gave score $\in$ [6;9] for valence and $\in$ [5;9] for arousal. 
This indicates that subjects tended to report scores around the average, and that apparently most of the interviews triggered a similar level of engagement. \added{The 10 subjects removed from the evaluation also show similar trends, with an average of 7.60 for arousal (standard deviation 1.91) and 7.59 for valence (standard deviation 1.93). Therefore, we argue that removing these subjects does not have a relevant impact on our descriptive statistics.}

\begin{table}[]
\caption{Descriptive statistics of the reported engagement.}
\label{tab:statisticsreports}
\centering
\resizebox{0.7\columnwidth}{!}{%
\begin{tabular}{ccccc}
\centering
 & Valence & Valence (norm) & Arousal & Arousal (norm) \\ \hline \hline
Average	& 7.23	& 1.90	& 7.06	& 2.13 \\ 
Minimum	& 1	& -4.94		&1	&-4.8 \\ 
Maximum	& 10	& 5.31		&10	&7 \\ 
Std. Dev.	&1.47	&1.58		&2.17	&2.17 \\ 
\end{tabular}
}
\end{table}

To gain more insight, it is useful to look at the reported engagement for each question. \added{In this case, we consider solely those subjects that responded to \textit{all} questions, i.e., 10 in total, to have a fair comparison}.  Figure~\ref{fig:valencearousalbyquestion} reports the box plots for valence and arousal for each question, divided by question group. We see that questions related to \textit{privacy}, \textit{ethics} and \textit{usage habits} tend to create more (positive) arousal in average, while questions related to \textit{procedures} are associated with more neutral values of arousal and valence (i.e., closer to 0 in the plot). Interestingly, questions related to \textit{relationships} show the largest variation in terms of arousal and valence (the box plot appears larger), indicating that this is a sensitive topic for the users, leading to more polarised scores in terms of emotional dimensions. The maximum average valence, instead, is observed for questions related to \textit{ethics}.

\subsection{RQ1: To  what  extent  can  we  predict  users’  reported engagement using biofeedback measurements and supervised  classifiers?}

In Table~\ref{tab:biofeedback} we report the performance of the different classifiers in terms of their precision, recall, F1-measure and accuracy, and considering the best configurations.  
Specifically, for each metric, we report the mean over the ten runs of the hold-out train-test procedure, i.e., the macro-average. This choice is in line with consolidated recommendations from literature on classification tasks using machine learning~\cite{Sebastiani}. Specifically, using macro-averaging is recommended with unbalanced data as ours, as it emphasizes the ability of a classifier to behave well also on categories with fewer training instances on specific classes. 
For each classifier, we report its best performance and the associated configuration options in terms of data balancing and scaling.


We see that the best performance (in \textbf{bold}) for valence and arousal are achieved by Random Forest (RF), when applying  balancing and standard scaling for valence (\textbf{Bal.} and \textbf{Scale} set to Y in the table), and balancing alone for arousal. The worst-performing algorithm is Naive Bayes (NB), regardless of the configuration. A general pattern cannot be identified about the influence of the different configuration options on the performance of the algorithms. However, for RF, we observe that compensation for class imbalance has a positive impact for valence, which is also the dimension characterised by a limited number of negative data points (see Table~\ref{tab:goldstandard}), thus confirming the effectiveness of SMOTE when class imbalance is present. 


\input{tables/table-biofeedback}


\input{tables/table-best-biofeedback}

In Table~\ref{tab:classificationResults}, we report the result of the two best algorithms with the best configurations, and we compare them with a baseline. Following previous research on sensor-based emotion recognition in software development~\cite{Girardi2020}, we select as baseline the trivial classifier always predicting the majority class, that is \textit{high} for arousal and \textit{positive} for valence. 
For the sake of completeness, we also report accuracy even if its usage is not recommended in the presence of unbalanced data as ours. 

For valence, the RF classifier distinguishes between negative and positive emotions with an F1 of $0.63$, thus obtaining an increment of $40$\% with respect to the baseline. Furthermore, we observe an improvement in precision of $58$\% (from $0.40$ of the baseline to $0.63$ of RF) and $28$\% in recall (from $0.50$ to $0.64$). These results indicate that the classifiers' behavior is substantially better than the baseline classifier that always predicts the positive class. 
  
As for arousal, we observe a comparable performance. The RF classifier distinguishes between high and low activation with an F1 of $0.65$, representing an improvement of $81$\% over the baseline ($0.36$). Again, the classifier substantially outperforms the baseline with an improvement of $135$\% for precision (from $0.28$ to $0.66$) and $0.32$\% for recall (from $0.50$ to $0.66$). The improvement with respect to the baseline is particularly high for arousal in terms of precision since arousal data (cf.  Table~\ref{tab:goldstandard}) are more balanced.  Therefore, the baseline that always predicts the positive class is inherently less effective, with a precision of $0.28$.




\subsection{RQ2: To what extent can we predict users’ engagement using voice analysis and supervised classifiers?}

Table~\ref{tab:tablevoiceresults} reports the comparison of the performance between the different machine learning algorithms, considering the best configurations. 
Differently from the biofeedback case, here we apply also data imputation (\textbf{Imp.} column) as a configuration option, by synthesising data for vectors with missing voice data, based on the corresponding biofeedback vectors. Therefore, the gold standard considered when Imp. is set to Y (Yes) is analogous to the one used for biofeedback and reported in Table~\ref{tab:goldstandard}. When Imp. is set to N (No), the gold standard is the one in Table~\ref{tab:goldstandardaudio}. 

\input{tables/table-voice}

This is an important aspect to notice for at least two reasons: (1) the two gold standards have slightly different distributions, and thus the default baselines for comparison will differ. In particular, when using imputation the baselines are the same as the one used in Table~\ref{tab:classificationResults}. When imputation is not used, the baselines need to be recomputed, and in particular the majority class baseline for arousal will always predict the low class, as this is the most frequent in Table~\ref{tab:goldstandardaudio}; (2) the usage of imputation in a real-world context assumes that, although solely voice data are used for classification, biofeedback data are collected anyway, so the practical advantage, both economic and logistic, is limited.    


\added{The best performance (in \textbf{bold}) is achieved by Support Vector Machines (SVM) for both arousal and valence, 
while NB and Decision Trees (DTree) remain far behind the other algorithms, regardless of the applied configuration options.} 

\added{
For valence, the best performance is achieved by SVM when balancing and scaling are applied and imputation is \textit{not} applied. The performance is higher than the ones obtained with biofeedback features (F1=$0.68$ for SVM vs F1=$0.63$ for RF, cf. Table~\ref{tab:biofeedback}).} 
This indicates that voice features alone appear to be particularly effective in discriminating the quality of the engagement (positive or negative valence), thereby confirming that our set-up is adequate for capturing the so-called \textit{emotional prosody}~\cite{buchanan2000recognition}, which reveals the sentiment of the speaker. Lower performance for valence is achieved with imputation for most of the algorithms (Table~\ref{tab:tablevoiceresults} reports the best-performing configurations, and Imp. = N for most of the high-performing cases). 
This suggests that biofeedback and voice capture different independent aspects of valence, as using biofeedback information to enrich voice data is not effective. 

\added{For arousal, the best performance is achieved by SVM when applying balancing and imputation, without scaling}. The performance is higher than the best one obtained with biofeedback features (F1=$0.71$ for SVM vs F1=$0.65$ for RF, cf. Table~\ref{tab:biofeedback}). \added{It should be noted that the application of imputation implies that one also leverages biofeedback features. If we consider SVM with balancing, scaling and no imputation, we however obtain F1=$0.69$ (results not reported in the table), which indicates that prediction can be effective also without resorting to biofeedback for imputation. Furthermore, while for biofeedback the best performing algorithm is RF, SVM achieves the best performance for voice.} 

\input{tables/table-best-voice}

To have additional insights, Table~\ref{tab:classificationResultsvoice} compares the result of the best algorithms, with respect to  the majority class baselines. 
\added{
For valence, we see an increase of $66$\% in terms of precision, $17$\% for recall, and $51$\% for F1. For arousal, the increase in performance is again higher, with $154$\% for precision, $42$\% for recall and $97$\% for F1. 
Let us now compare Table~\ref{tab:classificationResultsvoice} with Table~\ref{tab:classificationResults}, where we have the scores of the best-performing algorithm for biofeedback. 
We can see that for voice features are more effective than biofeedback ones not only in absolute numbers, but also compared to the majority class baselines (F1 increase is 51\% vs 40\% for valence, and 97\% vs 92\% for arousal). 
} 

Overall, our findings suggest that voice features represent a valid, and even more effective, alternative to biofeedback for emotion recognition during requirements elicitation. In fact, our classifiers' performance demonstrates that, in absence of biofeedback information, both valence and arousal can be successfully predicted with voice-only features.

\subsection{RQ3: To what extent can we predict users' engagement by combining voice and biofeedback features?}

Table~\ref{tab:tablecombine} reports the comparison of the performance of the different algorithms with their best configurations when using voice and biofeedback features combined. 

General trends are analogous to those observed when features are treated in separation, with the best performance achieved by RF and SVM, and lower performance by DTree and NB. 


\input{tables/table-combined}

\input{tables/table-best-combined}

\input{tables/table-best-of-best}

\added{
The best performance for valence (in \textbf{bold}) is achieved by RF, when applying SMOTE, and without scaling and imputation (F1=$0.68$). Instead, the best performance for arousal is obtained by SVM without SMOTE, and with scaling and imputation (F1=$0.69$). In Table~\ref{tab:classificationResultsCombined}, we compare the performance of the best algorithms with the majority class baselines. We see that for valence and arousal, we have an increase of 51\% and 92\%, respectively. However, it is important to notice that the combination of features does not lead to an increase in performance with respect to the case in which only voice features are used (cf. Table~\ref{tab:classificationResultsvoice} vs Table~\ref{tab:tablecombine}). While for valence the value of F1 is substantially the same (F1=$0.68$), for arousal the use of voice features only leads to the best performance (F1=$0.71$ vs F1=$0.69$). The same observations hold when comparing the results in terms of improvement with respect to the baselines. 
}

These observations are confirmed by Table~\ref{tab:best-of-best}, where the best performance of each algorithm is compared, considering different feature combinations. The table shows that the best performance in terms of F1 (in bold) is achieved either using voice features alone or, in some cases, by combining the features. 


%% file: tables/table-biofeedback.tex
\begin{table}[]
\centering
\caption{Performance of the different algorithms with the different configurations in terms of data balancing with SMOTE (Bal.) and in terms of standard scaling (Scale) when using \textbf{biofeedback} features. Y = Yes, the configuration option is applied; N = No, the configuration option is not applied.}
\label{tab:biofeedback}
\resizebox{0.5\columnwidth}{!}{
\begin{tabular}{c|c|c|c|c|c|c}
\hline
\multirow{3}{*}{\textit{Algorithm}} & \multicolumn{2}{c|}{\textit{Options}} & \multicolumn{4}{c}{\textit{Valence}} \\ 
\cline{2-7} 
 & \multirow{2}{*}{\textbf{Bal.}} & \multirow{2}{*}{\textbf{Scale}} & \multirow{2}{*}{\textbf{Prec}} & \multirow{2}{*}{\textbf{Rec}} & \multirow{2}{*}{\textbf{F1}} & \multirow{2}{*}{\textbf{Acc}} \\
 &  &  &  &  &  &  \\ \hline
SVM & Y & Y & 0.601 & 0.617 & 0.603 & 0.717 \\ 
MLP & N & N & 0.614 & 0.581 & 0.582 & 0.773 \\ 
DTree & N & Y & 0.576 & 0.569 & 0.569 & 0.731 \\ 
NB & N & Y & 0.550 & 0.525 & 0.497 & 0.773 \\ 
\textbf{RF} & Y & Y & \textbf{0.633} & \textbf{0.635} & \textbf{0.633} & \textbf{0.758} \\ \hline \hline
\multirow{3}{*}{\textit{Algorithm}} & \multicolumn{2}{c|}{\textit{Options}} & \multicolumn{4}{c}{\textit{Arousal}} \\ \cline{2-7} 
 & \multirow{2}{*}{\textbf{Bal.}} & \multirow{2}{*}{\textbf{Scale}} & \multirow{2}{*}{\textbf{Prec}} & \multirow{2}{*}{\textbf{Rec}} & \multirow{2}{*}{\textbf{F1}} & \multirow{2}{*}{\textbf{Acc}} \\
 &  &  &  &  &  &  \\ \hline
SVM & N & Y & 0.596 & 0.595 & 0.592 & 0.600 \\ 
MLP & Y & Y & 0.582 & 0.581 & 0.580 & 0.588 \\ 
DTree & N & N & 0.635 & 0.634 & 0.631 & 0.636 \\ 
NB & N & N & 0.514 & 0.510 & 0.468 & 0.544 \\ 
\textbf{RF} & Y & N & \textbf{0.658} & \textbf{0.656} & \textbf{0.654} & \textbf{0.660} \\ \hline
\end{tabular}
}

\end{table}

%% file: tables/table-best-biofeedback.tex
\begin{table}[h!]
\centering
\caption{Performance of the best classifiers based on F1, using EDA, BVP, and HR features with respect to majority class baseline classifier. Improvement over the baseline is also shown.}
\label{tab:classificationResults}
\resizebox{0.6\columnwidth}{!}{%
\begin{tabular}{l|llll}
 \textit{Algorithm} & \textbf{Precision} & \textbf{Recall} & \textbf{F1} & \textbf{Accuracy} \\ \hline
\multicolumn{5}{|c|}{\textit{Valence}} \\ \hline

\textit{Random Forest} & 0.63	& 0.64	& 0.63	& 0.76 \\
\textit{Baseline} & 0.40 & 0.50 & 0.45 & 0.79 \\
\textit{Improvement} & 0.23 (58\%) & 0.14 (28\%) & 0.18 (40\%) & -0.03 (-4\%) \\ 
\hline
\multicolumn{5}{|c|}{\textit{Arousal}} \\ \hline
\textit{Random Forest} & 0.66	& 0.66	& 0.65	& 0.66 \\
\textit{Baseline} & 0.28 & 0.50 & 0.36 & 0.56 \\
\textit{Improvement} & 0.38 (135\%) & 0.16 (32\%) & 0.29 (81\%) & 0.1 (18\%)
\end{tabular}%
}

\end{table}

%% file: tables/table-voice.tex
\begin{table}[]
\caption{\added{Performance of the different algorithms with the different configurations in terms of data balancing with SMOTE (Bal.), standard scaling (Scale) and Imputation (Imp.) when using \textbf{voice} features. Y = Yes, the configuration option is applied; N = No, the configuration option is not applied.}}
\label{tab:tablevoiceresults}
\centering
\resizebox{0.6\columnwidth}{!}{
\begin{tabular}{c|c|c|c|c|c|c|c}
\hline
\multirow{3}{*}{\textit{Algorithm}} & \multicolumn{3}{c|}{\textit{Options}} & \multicolumn{4}{c}{\textit{Valence}} \\ \cline{2-8} 
 & \multirow{2}{*}{\textbf{Bal.}} & \multirow{2}{*}{\textbf{Scale}} & \multirow{2}{*}{\textbf{Imp.}} & \multirow{2}{*}{\textbf{Prec}} & \multirow{2}{*}{\textbf{Rec}} & \multirow{2}{*}{\textbf{F1}} & \multirow{2}{*}{\textbf{Acc}} \\
 &  &  &  &  &  &  &  \\ \hline
\textbf{SVM}   & \textbf{Y} & \textbf{Y} & \textbf{N} & \textbf{0.680} & \textbf{0.674} & \textbf{0.675} & \textbf{0.805} \\
MLP   & Y & Y & N & 0.685 & 0.653 & 0.663 & 0.810 \\
DTree & Y & Y & N & 0.629 & 0.656 & 0.636 & 0.754 \\
NB    & N & Y & Y & 0.618 & 0.633 & 0.622 & 0.731 \\
RF   & Y & Y & N & 0.665 & 0.656 & 0.659 & 0.797

\\ \hline \hline
\multirow{3}{*}{\textit{Algorithm}} & \multicolumn{3}{c|}{\textit{Options}} & \multicolumn{4}{c}{\textit{Arousal}} \\ \cline{2-8} 
 & \multirow{2}{*}{\textbf{Bal.}} & \multirow{2}{*}{\textbf{Scale}} & \multirow{2}{*}{\textbf{Imp.}} & \multirow{2}{*}{\textbf{Prec}} & \multirow{2}{*}{\textbf{Rec}} & \multirow{2}{*}{\textbf{F1}} & \multirow{2}{*}{\textbf{Acc}} \\
 &  &  &  &  &  &  &  \\ \hline
\textbf{SVM}   & \textbf{Y} & \textbf{N} & \textbf{Y} & \textbf{0.715} & \textbf{0.717} & \textbf{0.713} & \textbf{0.716} \\
MLP   & N & Y & Y & 0.695 & 0.693 & 0.693 & 0.699 \\
DTree & Y & Y & N & 0.630 & 0.632 & 0.628 & 0.639 \\
NB    & N & N & Y & 0.634 & 0.636 & 0.632 & 0.635 \\
RF   & Y & N & Y & 0.688 & 0.688 & 0.686 & 0.691 \\
\hline
\end{tabular}
}

\end{table}

%% file: tables/table-best-voice.tex
\begin{table}[h!]

\caption{\added{Performance of the best classifiers, according to F1, using voice features with respect to majority class baseline classifier. Improvement over the baseline is also shown.}}

\label{tab:classificationResultsvoice} 
\centering
\resizebox{0.65\columnwidth}{!}{%
\begin{tabular}{l|llll}
 \textit{Algorithm} & \textbf{Precision} & \textbf{Recall} & \textbf{F1} & \textbf{Accuracy} \\ \hline
\multicolumn{5}{|c|}{\textit{Valence}} \\ \hline

\textit{SVM} & 0.68 & 0.67 & 0.68	& 0.81 \\
\textit{Baseline} & 0.41 & 0.50 & 0.45 & 0.81 \\
\textit{Improvement} & 0.27 (66\%) & 0.17 (34\%) & 0.23 (51\%) & 0.0 (0\%) \\ \hline

\multicolumn{5}{|c|}{\textit{Arousal}} \\ \hline
\textit{SVM} & 0.72	& 0.72	& 0.71	& 0.72 \\
\textit{Baseline} & 0.28 & 0.50 & 0.36 & 0.56 \\
\textit{Improvement} & 0.43 (154\%) & 0.21 (42\%) & 0.35 (97\%) & 0.16 (29\%)
\end{tabular}%
}

\end{table}


%% file: tables/table-combined.tex
\begin{table}[]
\caption{
\added{Performance of the classifiers with the different configurations, considering biofeedback and voice features \textbf{combined}.
}}
\label{tab:tablecombine}
\centering
\resizebox{0.65\columnwidth}{!}{
\begin{tabular}{c|c|c|c|c|c|c|c}
\hline
\multirow{3}{*}{\textit{Algorithm}} & \multicolumn{3}{c|}{\textit{Options}} & \multicolumn{4}{c}{\textit{Valence}} \\ \cline{2-8} 
 & \multirow{2}{*}{\textbf{Bal.}} & \multirow{2}{*}{\textbf{Scale}} & \multirow{2}{*}{\textbf{Imp.}} & \multirow{2}{*}{\textbf{Prec}} & \multirow{2}{*}{\textbf{Rec}} & \multirow{2}{*}{\textbf{F1}} & \multirow{2}{*}{\textbf{Acc}} \\
 &  &  &  &  &  &  &  \\ \hline
SVM   & Y & Y & Y & 0.657 & 0.647 & 0.650 & 0.779 \\
MLP   & Y & Y & N & 0.658 & 0.657 & 0.655 & 0.790 \\
DTree & Y & Y & N & 0.604 & 0.633 & 0.609 & 0.722 \\
NB    & Y & N & N & 0.613 & 0.645 & 0.617 & 0.726 \\
\textbf{RF}   & \textbf{Y} & \textbf{N} & \textbf{N} & \textbf{0.701} & \textbf{0.674} & \textbf{0.682} & \textbf{0.816}
\\ \hline \hline
\multirow{3}{*}{\textit{Algorithm}} & \multicolumn{3}{c|}{\textit{Options}} & \multicolumn{4}{c}{\textit{Arousal}} \\ \cline{2-8} 
 & \multirow{2}{*}{\textbf{Bal.}} & \multirow{2}{*}{\textbf{Scale}} & \multirow{2}{*}{\textbf{Imp.}} & \multirow{2}{*}{\textbf{Prec}} & \multirow{2}{*}{\textbf{Rec}} & \multirow{2}{*}{\textbf{F1}} & \multirow{2}{*}{\textbf{Acc}} \\
 &  &  &  &  &  &  &  \\ \hline
\textbf{SVM}   & \textbf{N} & \textbf{Y} & \textbf{Y} & \textbf{0.700} & \textbf{0.693} & \textbf{0.694} & \textbf{0.703} \\
MLP   & N & N & Y & 0.660 & 0.660 & 0.658 & 0.664 \\
DTree & Y & Y & Y & 0.655 & 0.654 & 0.649 & 0.653 \\
NB    & N & Y & Y & 0.620 & 0.621 & 0.618 & 0.621 \\
RF   & Y & N & Y & 0.694 & 0.692 & 0.690 & 0.695

\\ \hline
\end{tabular}
}

\end{table}

%% file: tables/table-best-combined.tex
\begin{table}[h!]

\caption{\added{Performance of the best classifiers, according to F1, using combined features with respect to majority class baseline classifiers. Improvement over the baseline is also shown.}}
\label{tab:classificationResultsCombined}
\centering
\resizebox{0.65\columnwidth}{!}{%
\begin{tabular}{l|llll}
 & \textbf{Precision} & \textbf{Recall} & \textbf{F1} & \textbf{Accuracy} \\ \hline
\multicolumn{5}{|c|}{\textit{Valence}} \\ \hline

\textit{Random Forest} & 0.70	& 0.67	& 0.68	& 0.82  \\
\textit{Baseline}  & 0.41 & 0.50 & 0.45 & 0.81 \\
\textit{Improvement} & 0.29  (71\%) & 0.22 (49\%) & 0.23 (51\%) & 0.1 (1\%) \\ \hline

\multicolumn{5}{|c|}{\textit{Arousal}} \\ \hline
\textit{SVM} &0.70 & 0.69 & 0.69	& 0.70  \\
\textit{Baseline} & 0.28 & 0.50 & 0.36 & 0.56 \\
\textit{Improvement} & 0.42 (150\%) & 0.19 (38\%) & 0.33 (92\%) & 0.14 (25\%)
\end{tabular}%
}

\end{table}

%% file: tables/table-best-of-best.tex
\begin{table}[]
\centering
\caption{\added{Comparison of the performance for all the algorithms, considering their best configurations for the different feature combinations. In \textbf{bold}, we report the best performance for each algorithm considering F1.}}
\label{tab:best-of-best}
\resizebox{0.8\columnwidth}{!}{
\begin{tabular}{c|cccc|cccc}
\hline
\multicolumn{1}{|c|}{\multirow{3}{*}{\textit{Algorithm}}} & \multicolumn{1}{c|}{\multirow{3}{*}{\textit{Features}}} & \multicolumn{3}{c|}{\textit{Valence}} & \multicolumn{1}{c|}{\multirow{3}{*}{\textit{Features}}} & \multicolumn{3}{c|}{\textit{Arousal}} \\ \cline{3-5} \cline{7-9} 
\multicolumn{1}{|c|}{} & \multicolumn{1}{c|}{} & \multicolumn{1}{c|}{\multirow{2}{*}{\textbf{Prec}}} & \multicolumn{1}{c|}{\multirow{2}{*}{\textbf{Rec}}} & \multicolumn{1}{c|}{\multirow{2}{*}{\textbf{F1}}} & \multicolumn{1}{c|}{} & \multicolumn{1}{c|}{\multirow{2}{*}{\textbf{Prec}}} & \multicolumn{1}{c|}{\multirow{2}{*}{\textbf{Rec}}} & \multicolumn{1}{c|}{\multirow{2}{*}{\textbf{F1}}} \\
\multicolumn{1}{|c|}{} & \multicolumn{1}{c|}{} & \multicolumn{1}{c|}{} & \multicolumn{1}{c|}{} & \multicolumn{1}{c|}{} & \multicolumn{1}{c|}{} & \multicolumn{1}{c|}{} & \multicolumn{1}{c|}{} & \multicolumn{1}{c|}{} \\ \hline

\multirow{3}{*}{SVM} & biofeedback & 0.601 & 0.617 & 0.603 & biofeedback & 0.596 & 0.595 & 0.592 \\
 & \textbf{voice} & \textbf{0.680}	& \textbf{0.674}	& \textbf{0.675} & \textbf{voice} & \textbf{0.715} & \textbf{0.717}	& \textbf{0.713}  \\
 & combined & 0.657	& 0.647	& 0.650  & combined & 0.700	& 0.693	& 0.694 \\ \hline

\multirow{3}{*}{MLP} & biofeedback & 0.580 & 0.583 & 0.578 & biofeedback & 0.582 & 0.581 & 0.580 \\
 & \textbf{voice} & \textbf{0.685}	& \textbf{0.653}	& \textbf{0.663}  & \textbf{voice} & \textbf{0.695} & \textbf{0.693}	& \textbf{0.693}  \\
 & combined & 0.658	& 0.657	& 0.655  & combined & 0.660	& 0.660	& 0.658  \\ \hline

\multirow{3}{*}{DTree} & biofeedback & 0.576 & 0.569 & 0.569 & biofeedback & 0.635 & 0.634 & 0.631 \\
 & \textbf{voice} & \textbf{0.629}	& \textbf{0.656}	& \textbf{0.636}  & voice & 0.630  & 0.632 & 0.628  \\
 & combined & 0.604	& 0.633	& 0.609  & \textbf{combined} & \textbf{0.655}	& \textbf{0.654}	& \textbf{0.649}  \\ \hline

\multirow{3}{*}{NB} & biofeedback & 0.550 & 0.525 & 0.497 & biofeedback & 0.514 & 0.510 & 0.468 \\
 & \textbf{voice} & \textbf{0.618}	& \textbf{0.633}	& \textbf{0.622}  & \textbf{voice} & \textbf{0.634} & \textbf{0.636} & \textbf{0.632} \\
 & combined & 0.613	& 0.645	& 0.617  & combined & 0.620	& 0.621	& 0.618  \\ \hline

\multirow{3}{*}{RF} & biofeedback & 0.633 & 0.635 & 0.633 & biofeedback & 0.658 & 0.656 & 0.654 \\
 & voice & 0.665	& 0.656	& 0.659  & voice & 0.688	& 0.688	& 0.686  \\
 & \textbf{combined} & \textbf{0.701}	& \textbf{0.674}	& \textbf{0.682}  & \textbf{combined} & \textbf{0.694}	& \textbf{0.692}	& \textbf{0.690}  \\ \hline
\end{tabular}
}

\end{table}

%% file: section/discussion.tex
\section{Discussion}
\label{sec:discussion}
The main take-away messages of this study are: 
\begin{itemize}
    \item product feedback interviews are activities that \added{are associated with positive engagement of the involved users};
    \item different levels of engagement are experienced depending on the topic of the question, with topics such as \textit{privacy}, \textit{ethics} and \textit{usage habits} \added{associated with} higher engagement, and \textit{relationships} \added{associated with} larger variations of engagement;
    \item by combining biofeedback features into vectors and by training the  Random Forest (RF) algorithm, it is possible to predict the engagement in a way that outperforms a majority-class baseline, with F1-measure of 63\% for valence and 65\% for arousal;
    \item using voice features only when training Support Vector Machines (SVM), performance is increased. Engagement can be predicted through voice features alone with F1-measure of 68\% (valence) and 71\% (arousal);
    \item \added{the combination of biofeedback and voice features does not increase the performance, with F1-measure 68\% (valence) and 69\% (arousal). In this case, the best performance is achieved with SVM for arousal, and with RF for valence;}
    \item \added{the results suggest that voice features are the best option to achieve effectiveness while ensuring greater affordability of the set-up and lower intrusiveness. In other terms, one can effectively predict engagement using voice feature only, i.e., using a simple microphone for recording, and without resorting to biofeedback, thus avoiding the possibly uncomfortable sensation of being monitored with a wristband.}
\end{itemize}

In the following sections, we discuss our results \rev{regarding engagement raised by the different interview topics}, \rev{compare them with} existing literature, and outline possible applications and timely avenues of research that are enabled by the current study.


\subsection{Engagement and Topics}
Our descriptive statistics  indicate that users experienced different levels of engagement with respect to the topic of a question.
Specifically, our participants reported a positive attitude when discussing privacy, ethics, and usage habits.
Concerning privacy and ethics, these topics were selected on purpose to trigger higher engagement.
Given the rising interest in these two fields, especially in relation to Facebook and online communities in general---e.g.,~\cite{trice2019values}---the obtained results are not surprising.  
Concerning \textit{usage habits}, we expected to see lower values of arousal.
Questions regarding usage habits were asked at the beginning.
Therefore, the observed high arousal could result from the excitement of the new experience.
However, we observed that question 19 (also about usage habits, but asked later) had the highest average arousal (3.6 in normalised values, while the average for usage habits questions is 2.8) and valence (3.2 vs. 2.5)\footnote{Results for each individual question not shown in the paper.}. 
We therefore argue that talking about usage habits triggers positive engagement. 
This indicates that users generally like the platform and are interested in sharing their relationship with it.
Qualitative analysis of the transcripts of the actual answers, not performed in this study, can further clarify these aspects. 
Overall, these results show that 1) users' interviews elicit emotions and engagement, with varying degrees of reactions depending on the topic; and 2) some topics are perceived as more engaging than others. \added{In our work, we decided to exclude neutral cases, to reduce the complexity of the problem of engagement prediction. This is also the choice made by Müller and Fritz~\cite{MF15}. However, neutral cases can also be important as they can be associated with features that are neither controversial nor exciting, and for which there is no particular need for improvement. In our case, most of the questions associated with neutral engagement in terms of valence and arousal are related to \textit{procedures}. This suggests that the application can be considered sufficiently easy to use and intuitive. Indeed, users recalling a specific procedure do not appear to encounter difficulties that may trigger particularly strong, possibly negative,  emotions.} 

\subsection{Performance Comparison with Related Studies}


According to the theoretical model of affect described in Sect.~\ref{sec:background}, in this study we use emotions as a proxy for engagement. 
Specifically, we operationalise emotions along the valence and arousal dimensions of the Circumplex Model of affect~\cite{Russel1991}, which we recognise using biofeedback and voice.
Using machine learning, we are able to classify the engagement of users during requirements interviews by distinguishing between positive and negative valence and high and low arousal. We experimented with different settings ---i.e., data balancing using SMOTE, data scaling, and data imputation (for voice data only). \added{Although the majority of the existing works in biometrics for software engineering do not focus on user feedback but rather on developers' emotions, it is useful to compare the obtained performance when similar sensors and sources of data are used.}

\paragraph{Biofeedback} As for our results using biofeedback features, 
a direct comparison is possible with the performance achieved in the empirical study by Girardi et al.~\cite{Girardi2020}, \added{who study the emotions of developers while programming in terms of valence and arousal}. The authors use the same device as ours (i.e., Empatica E4 wristband) and include the same features for EDA, BVP, and HR. 
Our classifier performance for arousal (F1=$0.65$) and valence (F1=$0.63$) outperforms the one they obtain---i.e., $0.55$ for arousal and $0.59$ for valence. Our better performance can be specifically linked to the task of interviewing, in which voice is not only used as a feature for predicting emotions, but also as means for expressing them~\cite{Laukka2017,Scherer2003}. The simple act of vocalizing can be regarded as an explicit, although not necessarily voluntary, expression of emotion.
This could affect other biometric aspects, improving the performance of our classifiers also when biofeedback-based features are the only predictors. \added{In other terms, the act of speaking can somewhat amplify the emotions of the subject, and this can affect the other biometric features sensed by a wristband.} 
Gerardi et al. report slightly better performance, though still lower than ours, when including also the signals gathered using an EEG helmet (F1=$0.59$ for arousal and F1=$0.60$ for valence). \added{In a more recent study, Girardi et al.~\cite{girardi2021emotions} consider the relation between emotions and perceived productivity in the workplace. They also use the Empatica E4 wristband, and the performance of their classifiers for valence reaches a maximum F1=$0.71$ for the best run and F1=$0.58$ as the average across runs (arousal is not evaluated in this study). Our average F1 for valence is $0.63$.}

Similarly to Girardi et al.~\cite{Girardi2020}, M{\"u}ller and Fritz~\cite{MF15} study \added{emotions while programming, focusing solely on valence}. They report an accuracy of $0.71$ \added{(our best accuracy using biofeedback is $0.76$)}, using a combination of features based on EEG, HR, and pupil size captured by an eye-tracker. \added{
In another work, Fritz et al.~\cite{FritzBMYZ14}, focus on the binary prediction of programming task's difficulty (i.e., easy vs difficult), and use EDA like in our study, plus eye-tracking and EEG as data sources, as well as different traditional machine learning algorithms for prediction. Their performance is comparable to ours, with the best F1=$0.73$. The study by Müller and Fritz~\cite{MullerF16} aims to predict code quality concerns (e.g., bugs, coding style violations) while a
developer changes the code. They again use Empatica E4, plus a SenseCore chest strap, to detect EDA and HR, and obtain a maximum precision and recall of   $0.22$ and $0.40$, respectively. Züger et al.~\cite{ZugerMMF18} conducted a field study to identify interruptibility of developers, i.e., whether it is acceptable to interrupt developers during their programming activity. They use a Fitbit Charge 2 wristband, a P7 chest strap, and they monitor their computer activities. Their best accuracy is $0.75$, when considering a combination of features. Finally, Fucci et al.~\cite{FucciGNQL19} target the identification of what kind of tasks developers are working on using EDA, HR, and EEG. The best accuracy ($0.87$) is achieved with HR-related features only. 
Although part of these numbers, especially the results Fucci et al.~\cite{FucciGNQL19}, may indicate that higher performance can be achieved using biometrics, the goals of all these studies are highly different from ours, as they deal with programming tasks. Besides that, some of the studies also use EEG, eye-tracking, or chest straps, which are either hardly applicable to face-to-face interviews (i.e., eye-tracking), or rather invasive (i.e., EEG, chest strap)}.

\paragraph{Voice} Considering our results when using voice features, our approach achieves higher performance with respect to biofeedback for arousal (F1$=0.71$) and valence (F1$= 0.68$).  
The results of our voice-based classifier are in line with a large part of the several machine learning-based state-of-the-art approaches in the broader field of speech emotion recognition \added{reported in the review by Akçay et al.~\cite{akccay2020speech}. For example, Zhao et al.~\cite{zhao2019exploring} use Long Short-Term Memory (LSTM) neural networks to predict emotions using voice spectrograms, achieving a maximum accuracy of $68.1\%$ (our best accuracy is $0.81$).
However, the availability of curated datasets in this field compared to our context has made it possible to achieve higher performance, even when considering larger sets of emotions, e.g., anger, sadness, fear/anxiety, neutral, happiness, disgust, and boredom. For instance, Issa et al.~\cite{issa2020speech} achieves an accuracy of up to $0.96$ when considering seven classes of emotions, and using convolutional neural networks (CNNs). Given these results, we argue that using existing datasets from the voice-based emotion recognition field to train machine learning classifiers, and then testing them on feedback interview data could improve the effectiveness of voice-based classifiers also for our context.}




\paragraph{Other Studies} As far as the combination of biofeedback and voice is concerned, our classifier shows comparable performance with respect to a deep-learning approach recently proposed by Aledhari et al.~\cite{AledhariRPS20}. As in our study, they use the Empatica E4 wristband for collecting biofeedback and report an accuracy of $0.85$ on the test set and $0.79$ in the validation set for the recognition of emotional valence, thus achieving performance comparable to ours ($0.82$ for combined features). \added{To our knowledge, this is one of the few studies using voice and biofeedback, although not related to software engineering activities.} Other studies in affective computing, also not related to software engineering activities, report much higher performance---e.g., accuracy of $0.97$ for arousal~\cite{Soleymani:TAFCC:SAM,Chen,Garcia} and $0.91$ for valence~\cite{nogueira2013hybrid}---for tasks in which emotions were recognized from standard stimuli (e.g., videos in the DEAP dataset~\cite{KoelstraMSLYEPNP12}).   
These studies rely on high-definition EEG helmets~\cite{Soleymani:TAFCC:SAM,Chen,Garcia} and facial electrodes for EMG \cite{nogueira2013hybrid}, which are invasive and cannot be used during real interviews with users or in remote interviews. 

\added{To summarise, the obtained results align with, or outperform, previous research in emotion recognition in software engineering using non-invasive technologies, i.e., F1 in the order of $0.6-0.7$, and accuracy around $0.8$. Higher accuracy is obtained only by studies using more invasive technologies, such as high-definition EEG helmets, or facial electrodes, which cannot be reasonably used in interviews. It is also important to remark that the other studies typically focus on developers’ emotions, e.g., while programming, rather than user or customer emotions during feedback interviews, so different affective states may be triggered by these diverse contexts.}

\subsection{Implications for Research and Practice} 
We consider this an exploratory study that helps us and the community to have a first understanding of engagement in user interviews and the potential usage of biofeedback devices and voice analysis in this context. \added{Since this is an exploratory study to assess the feasibility of the idea, we do not have a definite answer on whether the proposed approach for engagement prediction is sufficiently effective in practical settings. Furthermore, we cannot estimate the number of additional participants that would be needed to substantially increase the performance}. However, we argue that our results, once consolidated and confirmed by further studies, can have multiple applications and can open new avenues of research. 

\subsubsection{Applications in User Feedback} 
In user interviews similar to those staged in our experiment, biofeedback and voice information can be used to better investigate possible discrepancies between user engagement and the reported relevance of features. 
The ability to do so can facilitate requirements prioritization tasks the same way sentiment analysis does when applied to textual user feedback~\cite{Sut11}. \added{For example, by identifying discussion topics, or specific answers, that are associated with more positive engagement (e.g., relationships), requirements analysts can give priority to features that are related to those topics (e.g., identify friends or relatives in a social media app), or make these features more accessible from the graphical user interface. Conversely, identifying answers associated with negative engagement enables requirements analysts to understand what features need to be discarded or improved.}
Furthermore, the usage of these technologies can be extended to identify the engagement of the user \textit{on-the-fly}---i.e., during the interview---and help analysts steer the flow of the conversation. \added{For example, higher engagement may indicate that a certain topic is worth being investigated further, and the analyst can decide to ask more in-depth questions to discover new needs and possible features.}
These applications, which support human analysts in their activity, become particularly important when \textit{artificial agents} are used to elicit feedback or provide customer support, as shown by related research on voice analysis for call centers~\cite{han2020ordinal,li2019acoustic}. 
In these contexts, the detection of negative emotions is used to understand when a human operator needs to replace an artificial one because the latter is irritating the customer. 
\added{In our context, artificial agents, e.g., chatbots, can be used to replace human analysts, especially if one wants to collect feedback from a large user base automatically. While human analysts can, in principle, infer the degree of engagement during a conversation---especially if the interview is carried out face-to-face---a chatbot requires the support of emotion-aware tools such as those investigated in our study, to support automatic feature prioritisation and steer the conversation. It is worth noting that bots are an emerging area of research in software engineering~\cite{santhanam2022bots,9864620}, and proposals of chatbots have also been introduced for feedback analysis~\cite{wolfinger2022chatbot}. We argue that, given the technologies available today, e.g., based on the chat paradigm such as ChatGPT and similar tools, it will become more common, and acceptable for users to interact with chatbots to provide product feedback. This can also be done with a noninvasive approach by having the chatbot produce textual questions and having the user reply verbally. Interactive artificial intelligence, which is the usage of software to interact with people and automatically perform tasks---also thanks to the interaction with other software---is forecasted as a trend for the near future\footnote{\url{https://www.technologyreview.com/2023/09/15/1079624/deepmind-inflection-generative-ai-whats-next-mustafa-suleyman}. Visited 20 Sept. 2023.}.} 
Our work therefore opens to further applications on emotion-aware, voice-based chatbots for user interviews. 

\subsubsection{The Role of Voice} The introduction of voice features is particularly crucial. Biofeedback needs to be locally acquired with specialised devices such as Empatica E4, which (i) costs about \$1,690.00 at the time of writing; (ii) needs to locally register the different signals; (iii) does not remotely send the signal in an automated manner; (iv) can raise privacy concerns. Therefore, their usage is realistic only during face-to-face interviews, in which a certain level of mutual trust can be achieved and all data can be acquired locally. Instead, the analysis of voice is particularly appropriate in remote communication  scenarios---involving either human or artificial agents---which are increasingly common due to the COVID-19 pandemic. Voice is voluntarily produced and transmitted by users, and can be remotely recorded and processed without resorting to specialised devices, with evident cost savings. The cost reduction extends the applicability of the idea to large-scale scenarios. \added{With voice analysis, automated user feedback campaigns become feasible, and companies can improve automated A/B testing of web apps or pages. Specifically, they can ask multiple users to interact with different versions of an interface, and speak up their reflections on the experience. The recording and the analysis of the engagement can be used to facilitate the identification of preferred versions, appreciated features, or relevant interaction problems.} 

\subsubsection{Applications in RE and Software Engineering.} In the case of more classical requirements elicitation interviews~\cite{davis2006effectiveness,zowghi2005requirements}, the usage of biometrics can support these activities to improve the analyst's ability to create a trustworthy relationship \added{with} the customer, and improve the quality of the interview and the collected data. 
In this context, it is relevant to extend the work to identify the customer's frustration, which often corresponds to the first step to create mistrust in the analyst~\cite{distanont2012engagement}.
Frustration can be detected using biofeedback by analyzing the changes in the heart rate, temperature, and other vitals~\cite{Haag2004, citeulike:2863490, 159-mandryk_bit2006,Scheirer:2002p4107} and used to warn the analyst. Furthermore, frustration is strictly related with stress, which can be detected in voice signals through Teager energy operator (TEO)-based features~\cite{zhou2001nonlinear,bandela2017stressed}.

Overall, we argue that the analysis of voice, with its relative cost-effecti\-ve\-ness, can be broadly applied not only to RE, but to all software engineering scenarios in which conversations are central (e.g., SCRUM stand-up meetings, synchronous code reviews, pair-programming~\cite{ZP21}) to investigate the emotional side of these human-intensive activities that have a relevant impact on the development, but are currently ephemeral.

\subsubsection{Tacit Knowledge} It is worth noting that the improved performance obtained with voice features, and the lower cost of the approach, do not rule out biofeedback. Indeed, biofeedback captures involuntary body signals that the speaker cannot fully control, while voice tone can, to a certain extent, be manipulated to deceive~\cite{KimAndre}. Biofeedback can reveal a more faithful representation of emotions, and one can compare discrepancies between emotion prediction with biofeedback and with voice to identify situations in which what the voice ``tells'' is different from what the speaker ``feels.'' This can happen in requirements elicitation interviews, which can involve controversial political aspects~\cite{milne2012power}, or domain experts who need to be interviewed to gather process-related information but may be reluctant to share their knowledge~\cite{gervasi2013unpacking}. Therefore, the results of this research can be further applied to tackle the issue of tacit knowledge in requirements engineering~\cite{gervasi2013unpacking,ferrari2016ambiguitycues,sutcliffe2020known}. 




%% file: section/threats.tex
\section{Threats to Validity}
\label{sec:threats}

In this section, we report the main limitations of our study and how we address them.
The order in which we report the threats follows Wohlin et al.~\cite{WRH12} suggestions regarding exploratory research.

\textit{Internal validity}.  Threats to internal validity deal with confounding factors that can influence the results of a study.
We collected data in a laboratory setting, and factors existing in our settings, such as the presence of the experimenter, can influence the emotional status of the participants \rev{(i.e., the Hawthorne effect)}~\cite{adair1984hawthorne}.
Establishing a trust-based rapport with the participants in a relaxed setting is crucial to mitigate these threats.
Thus, we invited the participant to wear the wristband when entering the room, before the actual interview started, in order to get acquainted with the device, settings, and the presence of the experimenter.
Furthermore, self-assessment questionnaires were filled \textit{immediately after} the interview.
This choice was driven by the need to preserve a realistic interview context.
However, with this design, the engagement is \textit{recalled} by the subject and not reported in the moment in which it emerged.
Therefore, discrepancies can occur between the feeling of engagement and its rationally-processed memory.
Similarly, to keep a realistic setting, we did not perform pre-interviews to assess the participants' mood (i.e., the presence of a long-lasting emotion) or their personality traits. 
We acknowledge that an emotionally-charged event in the life of a participant, occurring before the interview, can impact the results. 

\added{The designed interview script covers a wide variety of topics, also possibly controversial ones (e.g., questions 16, 17, and 34 to 38), but we observed a generally positive degree of engagement. This can be because the feedback interview was actually fictional, and the participants were aware that the activity was part of a research project. Therefore, their reactions could be due to their willingness to please the researcher. In addition, it could be hard to trigger a strong reaction if a user is not actually using a product. A possible way to address this could be to use an actual version or mock-up of the app during the interview, and let the user interact with it to elicit better, e.g., frustration or other strongly negative emotions.}

In our design, we consider a \textit{hold-out} setting for validation in line with previous studies~\cite{girardi2021emotions,Girardi:2019,MullerF16}---i.e., a subject could contribute with different vectors to training and test sets; however, in practice, this can happen only if the same subject is interviewed multiple times. 
This choice is driven by the limited number of subjects, which does not allow to have effective predictions in a more realistic \textit{leave-one-subject-out} (LOSO) setting.
For example, Girardi et al.~\cite{girardi2021emotions} perform an \textit{in vivo} study involving 21 developers using the LOSO evaluation settings, showing an F1=$0.46$ for valence.

\textit{Construct validity}. This threat refers to the reliability of the operationalization of the study constructs.
\added{Our study may suffer from threats to construct validity in capturing emotions using self-reports. Self-reported engagement may differ from the actual engagement, and this threat is however unavoidable as it depends on the honesty and the self-reflection ability of the participants. Furthermore, some participants may not report any actual variation of engagement, thereby making their feedback not relevant for measuring differences in terms of this construct.}
To address this issue, we performed data quality assurance and excluded participants who did not show engagement with the task (e.g., who provided always the same score, or scores with overall low standard deviation). 
We believe that the designed interview script is sufficiently representative of typical users' interviews in terms of triggered engagement. 
However, we did not pause the interview when addressing a new topic without giving the respondents the chance to relax and reduce their emotional agitation (e.g., by watching a relaxing video). 
Therefore, the order in which we presented the topics can have an impact on the result---i.e., for some topics presented later in the interview we may be observing an \textit{emotional spillover} due to emotional charge introduced when discussing earlier topics. Concerning the self-assessment questionnaire, this was adapted from previous studies in software engineering~\cite{MullerF16}, and the users gained confidence with self-assessment in the initial emotion-triggering activity. 

We exclude neutral cases from our dataset.
This choice is driven by the need to simplify the prediction problem.
Considering the exploratory nature of the study,  we decided to focus first on more polarised emotional states, which are assumed to be clearly distinguishable.
Preliminary experiments including neutral cases showed an F1=$0.45$ for both valence and arousal using RF with biofeedback features, in a hold-out setting like the one adopted in the rest of the experiments of our paper.

\textit{Conclusion validity}. 
The validity of our conclusions relies on the robustness of the  machine learning models.
To mitigate any threat arising from having a small dataset, we ran several algorithms addressing the same classification task.
In all runs, we performed hyperparameters tuning as recommended by state-of-the-art research~\cite{tantithamthavorn2018impact}.
Following consolidated guidelines for machine learning, we split our data into train-test subsets.
The training is performed using cross-validation and the final model performance is assessed on a hold-out test set. The entire process is repeated ten times for each algorithm, to account for random variations in the data.
Moreover, our classifiers configuration included scaling and data balancing techniques. \added{Although the usage of different configurations may lead to question the robustness of the proposed idea, we remark that this is a feasibility study, and exploring different settings is part of the exploratory nature of this type of research.}

\textit{External validity}.  
The generalizability of our results is limited by the number of subjects (and associated data points) who took part in the study. \added{The limited data may have also impacted the performance.} \added{Given the exploratory nature of the study, we argue that the number of participants can however be considered acceptable, as it is comparable with similar contributions~\cite{MF15,vrzakova2020affect,Girardi2020,FucciGNQL19}, for which the number of participants ranges from 17~\cite{MF15} to 33~\cite{vrzakova2020affect}.  Products such as Affectiva for emotion recognition from videos use a dataset of about 6 million faces, from 38,944 hours of data\footnote{\url{https://blog.affectiva.com/the-worlds-largest-emotion-database-5.3-million-faces-and-counting}. Visited 19 Sept. 2023.}, which gives an indication of the number of data points needed to develop a commercial product.}
Although with some imbalance, our sample includes multiple ethnic groups and genders to account for physiological differentiation~\cite{bent2020investigating}. 
Further replications with a confirmatory design should engage more participants, and consider balance between ethnicity, culture, age, and gender to account for the differences in emotional reactions due to these aspects. 
\rev{Handling of physiological data, such as the ones used in this study, can raise ethical concerns once our proposed approach is transferred to real-world settings. This is an ecological threat to the validity of our study since potential stakeholders may not take part in or hold back during the interviews.}

As for the topic of the interviews, we selected features from a commonly-used social media app for which no particular expertise is needed. 
We explicitly base our questionnaire on the Facebook social media platform due to its familiarity for both interviewers and interviewees. However, the questions are generic enough to cover functionalities of other platforms in the same space. 
The interviews were performed in 2016 before the Facebook–Cambridge Analytica data scandal was reported by \textit{The New York Times} in December 2016~\cite{Funk2016NYT}. Up to that point, the only report of profiles harvested on Facebook was published by the Guardian in December 2015~\cite{Davies2015Guardian}, and Facebook was not considered a controversial platform.  
\added{It is worth noting that the perception of social media apps may have changed in recent years, also due to the Cambridge Analytica scandal. In case the experiment is replicated, e.g., on Facebook or a similar app, we may identify different emotions triggered by the same questions. This affects our observations about the relationships between specific topics (e.g., ethics, usage habits) and engagement (cf. Section~\ref{sec:descriptivestatistics}). However, the most relevant outcome of our study, associated with our research questions, is that emotions can be predicted with machine learning methods, using biofeedback and voice features. We argue that this conclusion will not change if the experiment is replicated in a more contemporary context.}




%% file: section/conclusion.tex
\section{Conclusion and Future Work}
\label{sec:conclusion}

This paper presents the first study about engagement prediction in product feedback interviews. In particular, we show that it is possible to predict the positive or negative engagement of a user during an interview about a product. This can be achieved through the usage of \textit{biofeedback} measurements acquired through a wristband, the analysis of \textit{voice} through audio processing, and the application of supervised machine learning. 
In particular, we show that voice analysis alone can lead to sufficiently good results.

The study is exploratory in nature, and application of our results requires further investigation, and the acceptance of the non-intrusive, yet potentially undesired, biofeedback device.
Nevertheless, we believe that the current work, with its promising results, establishes the basis for further research on engagement, and emotions in general, during the many human-intensive activities of system development. 
Among the future works, we plan to: (a) replicate the experiment with a larger and more representative sample of participants, utilizing a realistic validation setting---e.g., including neutral data; (b) complement our analysis with the usage of other emotion-revealing signals, such as facial expressions captured through cameras~\cite{SoleymaniAFP16} and electroencephalographic (EEG) activity data~\cite{Girardi2020,MF15}; (c) tailor the study protocol for requirements elicitation interviews for novel products yet to be developed; (d) analyse emotions of interviewers and interviewee during the dialogue, to investigate if it is possible to identify the process of construction of trust;   
(e) investigate and compare the emotional footprint of different software development-related tasks. This can be done for example by looking at the difference between physiological signals of the multiple actors of the development process across different phases, such as of development, elicitation, testing, \textit{etc}; (f) apply voice analysis to SCRUM meetings, focus groups, and other software engineering activities in which speech plays a primary role. 
